\documentclass[paper]{JHEP3} 
\usepackage{amsmath}
\usepackage{amssymb}
\usepackage{graphicx}
\def\be{\begin{equation}}
\def\ee{\end{equation}}
\def\bea{\begin{eqnarray}}
\def\eea{\end{eqnarray}}
\def\rarr{\rightarrow}
\def\nn{\nonumber}
\def\tr{\mbox{tr}\,}

\def\rarr{\rightarrow}

\def\kf{{\bf k}}
\def\qf{{\bf q}}
\def\lf{{\bf l}}

\newcommand{\mbf}[1]{\mbox{\boldmath $#1$}}
\newcommand{\bl}{\lf}

\newcommand{\bN}{\mbf{N}}
\newcommand{\bk}{\kf}
\newcommand{\bq}{\qf}
\newcommand{\beps}{\mbf{\epsilon}}

\newcommand{\cL}{{\cal L}}
\newcommand{\lambdab}{\bar{\lambda}}

\newcommand\C{{\rm\kern.24em
    \vrule width.02em height1.4ex depth-.05ex
    \kern-.26em C}}

\def\Db{{\mathchoice{\hbox{$\displaystyle\mathbb D$}}
        {\hbox{$\textstyle\mathbb D$}}{\hbox{$\scriptstyle\mathbb D$}}
        {\hbox{$\scriptscriptstyle\mathbb D$}}}}
\def\Bb{{\mathchoice{\hbox{$\displaystyle\mathbb B$}}
        {\hbox{$\textstyle\mathbb B$}}{\hbox{$\scriptstyle\mathbb B$}}
        {\hbox{$\scriptscriptstyle\mathbb B$}}}}

\def\nn{\nonumber}
\def\fr{\frac}

\def\rarr{\rightarrow}

\newdimen\picraise
\newcommand\picb[1]
{
  \setbox0=\hbox{\includegraphics{#1}}
  \picraise=-0.5\ht0
  \advance\picraise by 0.5\dp0
  \advance\picraise by 3pt     % correct for height of `=' above baseline
  \hbox{\raise\picraise \box0}
}

\newdimen\picraise
\newcommand\picresize[2]
{
  \setbox0=\hbox{\includegraphics[width=#2]{#1}}
  \picraise=-0.5\ht0
  \advance\picraise by 0.5\dp0
  \advance\picraise by 3pt     % correct for height of `=' above baseline
  \hbox{\raise\picraise \box0}
}

\newdimen\picraiseano
\newcommand\eqeps[2]
{
  \setbox0=\hbox{\includegraphics{#1}}
  \picraiseano=-0.5\ht0 
  \advance\picraiseano by 0.5\dp0
  \advance\picraiseano by -#2pt  % correct for height of pictures in inteq. 
  \hbox{\raise\picraiseano \box0}
}

\newdimen\picraiseano
\newcommand\eqepsres[3]
{
  \setbox0=\hbox{\includegraphics[width=#3]{#1}}
  \picraiseano=-0.5\ht0
  \advance\picraiseano by 0.5\dp0
  \advance\picraiseano by -#2pt  % correct for height of pictures in inteq.
  \hbox{\raise\picraiseano \box0}
}

\newdimen\picraise
\newcommand\picbox[1]
{
  \setbox0=\hbox{\input{#1}}
  \picraise=-0.5\ht0 
  \advance\picraise by 0.5\dp0
  \advance\picraise by 3pt     % correct for height of `=' above baseline
  \hbox{\raise\picraise \box0}
}

\newdimen\picraiset
\newcommand\picding[1]
{
  \setbox0=\hbox{\input{#1}}
  \picraiset=-0.5\ht0 
  \advance\picraiset by 0.5\dp0
  \advance\picraiset by -3pt  % correct for height of pictures in inteq. 
  \hbox{\raise\picraiset \box0}
}

\newdimen\picraisehallo

\newcommand\pichallo[2]
{
  \setbox0=\hbox{\input{#1}}
  \picraisehallo=-0.5\ht0 
  \advance\picraisehallo by 0.5\dp0
  \advance\picraisehallo by -#2pt  % correct for height of pictures in inteq. 
  \hbox{\raise\picraisehallo \box0}
}

\preprint{DESY-09-200\\
HD-THEP-08-11\\
BI-TP 2009/27\\
ECT$^*$-07-07
}
\title{
\boldmath 
High Energy Behavior of a Six-Point $R$-Current Correlator 
in ${\cal N}=4$ Supersymmetric Yang-Mills Theory
\unboldmath}
\author{Jochen Bartels\hspace{1pt}${}^{a,1}$, 
Carlo Ewerz\hspace{1pt}${}^{b,c,d,e,2}$, 
Martin Hentschinski\hspace{1pt}${}^{a,3}$, 
Anna-Maria Mischler\hspace{1pt}${}^{a,4}$
\\
$^a$
II.\ Institut f\"ur Theoretische Physik, Universit\"at Hamburg,\\
\phantom{$^a$}
Luruper Chaussee 149, D-22761 Hamburg, Germany
\\
$^b$ 
Institut f\"ur Theoretische Physik, Universit\"at Heidelberg,\\
\phantom{$^d$}
Philosophenweg 16, D-69120 Heidelberg, Germany\\
$^c$
ExtreMe Matter Institute EMMI, GSI Helmholtzzentrum f\"ur Schwerionenforschung,\\
\phantom{$^b$} 
Planckstra{\ss}e 1, D-64291 Darmstadt, Germany\\
$^d$
Fakult\"at f\"ur Physik, Universit\"at Bielefeld, D-33615 Bielefeld, Germany
\\
$^e$
ECT\,$^*$\!, Strada delle Tabarelle 286, 
I-38050 Villazzano (Trento), Italy
\\
\\
$^1$E-mail: \email{Jochen.Bartels@desy.de}\\
$^2$E-mail: \email{C.Ewerz@thphys.uni-heidelberg.de}\\
$^3$E-mail: \email{Martin.Hentschinski@desy.de}\\
$^4$E-mail: \email{Anna-Maria.Mischler@desy.de}\\
} 
\abstract{
We study the high energy limit of a six-point $R$-current
correlator in ${\cal N}=4$ \linebreak supersymmetric Yang-Mills
theory for finite $N_c$. We make use of the framework of
perturbative resummation of large logarithms of the energy. More
specifically, we apply the (extended) generalized leading
logarithmic approximation. We find that the same conformally
invariant two-to-four gluon vertex occurs as in non-supersymmetric
Yang-Mills theory. As a new feature we find a direct coupling of
the four-gluon $t$-channel state to the $R$-current impact factor.}


\begin{document}

\section{Introduction}
\label{sec:intro}

The high energy limit of nonabelian gauge theories, in particular of
QCD, has been extensively studied in a variety of perturbative and
nonperturbative approaches. In the groundbreaking work of
\cite{Kuraev:fs,Balitsky:ic} the perturbative resummation of leading
logarithms of the center-of-mass energy $\sqrt{s}$ was performed. The
resulting leading logarithmic approximation (LLA) is encoded in the
celebrated BFKL equation. It collects all perturbative terms of the
order $(\alpha_s \log s)^m$ in which the smallness of the strong
coupling constant $\alpha_s$ is compensated by large logarithms of the
energy. For the scattering amplitude of $2 \to 2$ scattering
processes, this leading logarithmic approximation corresponds to
resumming diagrams in which two interacting reggeized gluons are
exchanged in the $t$-channel. This bound state of two gluons, the
Pomeron, can hence be represented diagrammatically by a gluon ladder.

Based on Gribov's work on the Reggeon calculus \cite{Gribov:1968fc} it
was immediately clear that, once such moving Regge singularities
exist, the high energy behavior of nonabelian gauge theories can be
formulated in terms of an effective 2+1 dimensional field theory,
called Reggeon field theory. It lives in the two transverse dimensions
of the scattering process, and rapidity can be understood as a
timelike parameter. Steps towards explicitly formulating this
effective high energy description include the generalization of the
BFKL approximation to the evolution of $n$-gluon states, known as the
BKP equations \cite{Bartels:1980pe,Kwiecinski:1980wb}, and a further
generalization which encompasses number-changing processes during the
$t$-channel evolution. The latter is known as the (extended)
generalized leading logarithmic approximation, (E)GLLA, and will be
described in some detail later in this paper. It has been used to
derive a $2\to 4$ gluon vertex that contains the triple Pomeron vertex
\cite{Bartels:1992ym,Bartels:1993ih,Bartels:1994jj}, as well as a
higher order $2\to6$ gluon vertex function \cite{Bartels:1999aw}. A
systematic approach of deriving, for nonabelian gauge theories, the
elements of Reggeon field theory, is the effective action developed in
\cite{Lipatov:1991nf,Lipatov:1995pn}, see \cite{Hentschinski:2009cc}.
Other approaches to the problem of understanding the high energy limit
of QCD include the Wilson line operator expansion
\cite{Balitsky:1995ub}-\cite{Balitsky:2001mr}, the dipole picture of
high energy scattering \cite{Mueller:1993rr}-\cite{Chen:1995pa}, and
the color glass condensate approach, see for example
\cite{Iancu:2003xm}. In the limit $N_c \to \infty$ the latter three
approaches as well as the EGLLA all give rise to the same non-linear
evolution equation, known as the BK equation
\cite{Balitsky:1995ub,Kovchegov:1996ty,Kovchegov:1997pc}. All
approaches mentioned here are of perturbative nature. A
nonperturbative derivation of a Reggeon field theory from QCD still
appears prohibitively difficult.

Among the most remarkable features of Reggeon field theory for QCD are
the conformal invariance in the two-dimensional transverse coordinate
space \cite{Lipatov:1985uk,Bartels:1995kf,Ewerz:2001uq}, and the
integrability in the large-$N_c$ limit
\cite{Lipatov:1993yb,Lipatov:1994xy,Faddeev:1994zg}. Both properties,
so far, have been established only for the leading order of the
resummation in the LLA and (E)GLLA. A full understanding of the fate
of these symmetries in next-to-leading order is still missing.

With the advent of the AdS/CFT correspondence, which relates the
maximally supersymmetric nonabelian gauge theory in four dimensions,
that is ${\cal N}=4$ super Yang-Mills theory (SYM) with gauge group
$\text{SU}(N_c)$, to type IIB superstring theory on a five-dimensional
AdS space \cite{Maldacena:1997re,Gubser:1998bc,Witten:1998qj}, the
natural question appears whether Reggeon field theory has a dual
analog on the string (or supergravity) side. As a first step, one
identifies scattering amplitudes or correlators which are defined on
both sides of the correspondence: in QCD, a clean environment for
studying the dynamics at high energies has been found in $\gamma^*
\gamma^*$ scattering, i.\,e.\ in the four-point correlators of
electromagnetic currents. In ${\cal N}=4$ SYM, it has been suggested
\cite{CaronHuot:2006te} to consider, as a substitute for the
$\mbox{U}(1)$ current of electromagnetism, the $R$-currents which
result from the global $\text{SU}_R(4)$ $R$-symmetry. One is thus led
to investigate, in suitable high energy limits, correlators of
$R$-currents, both for ${\cal N}=4$ SYM and for the dual string
theory.

On the gauge theory side, existing QCD calculations provide a natural
starting point for a systematic investigation of this
correspondence. It is, however, clear that certain differences exist
between (non-supersymmetric) QCD and ${\cal N}=4$ SYM, and one has to
study their consequences. For the high energy behavior of current
correlators, it is the impact factors which are sensitive to
supersymmetry: in QCD, the fermions (quarks) belong to the fundamental
representation of the gauge group, whereas in ${\cal N}=4$ SYM all
particles (gluons, Weyl fermions and scalars) are in the adjoint
representation of the gauge group. On the other hand, since in quantum
field theory the high energy behavior is dominated by the exchange of
particles with the highest spin -- that is the gluons in both theories
-- the leading logarithmic approximation in ${\cal N}=4$ SYM should be
quite similar to QCD.

As the very first step, the elastic scattering of two $R$-currents in
${\cal N}=4$ SYM has been studied in \cite{Bartels:2008zy}. Apart from
the new impact factors, it has been verified that the high energy
behavior is dominated by the familiar BFKL Pomeron. Turning to
elements of Reggeon field theory beyond the BFKL two-gluon ladders,
the effect of supersymmetry is expected to be more severe. As a
theoretical environment for extracting the triple Pomeron vertex
calculations on the QCD side have made use of six-point
functions. This motivates, for the extension to ${\cal N}=4$ SYM, to
investigate six-point correlators of $R$-currents. It is the purpose
of the present paper to study the high energy behavior of such
$R$-current correlators in the extended generalized leading
logarithmic approximation. As a result we will find that ${\cal N}=4$
SYM provides a new element in Reggeon field theory which is not
present in non-supersymmetric QCD. On the other hand, the triple
Pomeron vertex remains the same as in QCD. We emphasize that
throughout this paper we keep $N_c$ finite, and we only briefly
comment on the large-$N_c$ limit at the end.

Parallel to this investigation of the gauge theory side, it is 
interesting to study the $R$-current correlators in the same kinematic 
limit also on the string side. Here, in the simplest approximation, 
one considers the zero slope limit and arrives at Witten diagrams with 
graviton exchanges. For the elastic scattering this has been done 
in \cite{Bartels:2009sc}, while for the problem of the six-point function 
work along these lines is in progress. 

Our paper is organized as follows. In section \ref{sec:sixpoint} we
define the triple-Regge limit of six-point correlators, first for
virtual photons in QCD, then for $R$-currents in ${\cal N}=4$ SYM
theory. In section \ref{sec:external} we consider the $R$-current
impact factors, consisting of the sum of a Weyl fermion loop and a
scalar loop in the adjoint representation with $n$ gluons attached.
We derive the relation of these impact factors to the corresponding
impact factors consisting of quark loops in the fundamental
representation, and present them explicitly for up to four gluons. We
put special emphasis on the reggeization of the impact factors.
Further in that section we point out that the Odderon decouples from
impact factors containing particles in the adjoint representation. In
section \ref{sec:inteq} we write down the integral equations which sum
all diagrams contributing to the (extended) generalized leading
logarithmic approximation. In section \ref{sec:amplitudeseglla} we
study these integral equations, tracing in particular the consequences
of the new impact factors obtained before. In section
\ref{sec:partialwave} we finally present our result for the six-point
correlator which differs from the result in QCD. Section
\ref{sec:outlook} contains our conclusions and an outlook. Appendix
\ref{sec:color} deals with some $\text{su}(N_c)$ color identities. In
two further appendices we consider higher $n$-gluon amplitudes and
their reggeization in ${\cal N}=4$ SYM: In appendix \ref{appfive} we
generalize our findings for the four-gluon amplitude to five gluons.
In appendix \ref{appsix}, finally, we make some steps towards a
calculation of the six-gluon amplitude. These first steps already
allow us to draw some conclusions about the 2-to-6 gluon transition
function in ${\cal N}=4$ SYM. Some results presented in this paper
have been published in a letter \cite{Bartels:2009ms}.

\section{Six-point correlation functions at high energies}
\label{sec:sixpoint}

In this section we define the high energy limit of six-point functions
in ${\cal N}=4$ SYM. As is well known, the high energy behavior of
scattering amplitudes in the Regge limit is determined by the exchange
of particles with the highest spin which, in the case of ${\cal N}=4$
SYM, are the gauge bosons with spin 1. Studies of the high energy
behavior of Yang-Mills theories in the leading logarithmic
approximation show that, apart from the impact factors, the high
energy behavior is entirely determined by the gauge bosons. This
implies that, in this approximation, in a supersymmetric extension of
non-supersymmetric $\text{SU}(N_c)$ gauge theory, e.\,g.\ in 
${\cal N}=4$ SYM with the same gauge group, the difference between the
supersymmetric and the non-supersymmetric theory resides in the impact
factors. Specifically, in supersymmetric theories the fermion fields
are in the adjoint representation, and in addition to the impact
factors consisting of closed fermion loops we have also those composed
of scalar particles. Apart from the impact factors, the interactions
of the exchanged gluons are the same as in the non-supersymmetric
case.

\subsection{Six-point amplitudes in QCD}

In QCD, scattering amplitudes with more than $4$ external particles
arise naturally in the context of deep inelastic scattering on a
weakly bound nucleus. A simple example is deep inelastic scattering
(DIS) on a nucleus consisting of two weakly bound nucleons (that is a
deuteron), see figure~\ref{fig:gammapn}.%
\FIGURE{
\includegraphics[width=8.6cm]{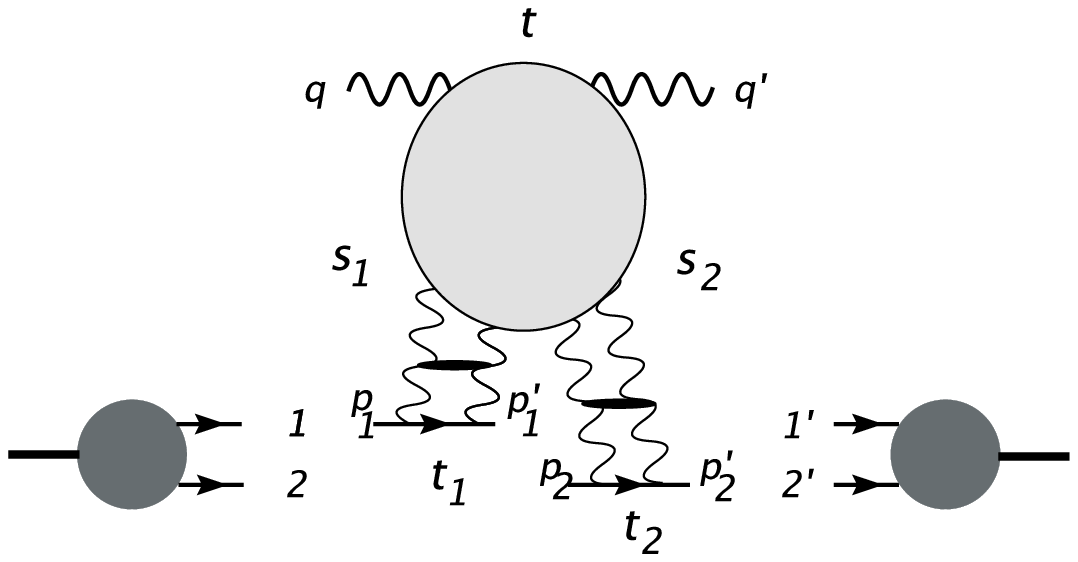}
\caption{Scattering of a virtual photon on a weakly bound nucleus
\label{fig:gammapn}}
}
The total cross section of this scattering process is obtained 
from the elastic scattering amplitude, $T_{\gamma^*(pn) \to \gamma ^* (pn)}$, 
via the optical theorem, 
\begin{equation}
\label{eq:sigmatot}
\sigma^{\rm tot}_{\gamma^*(pn) \to \gamma ^* (pn)} = \frac{1}{S} \,{\rm Im}\, 
T_{\gamma^*(pn) \to \gamma ^* (pn)} \,, 
\end{equation} 
where $S = (q + p_1 + p_2)^2$ denotes the total energy of the
scattering process. In order to obtain an entirely perturbative
environment, we can think of replacing the two nucleons by virtual
photons. As a result, we are led to six-point correlators of
(off-shell) electromagnetic currents, 
$T_{\gamma^* \gamma^* \gamma^*\to \gamma ^* \gamma^* \gamma^*}$, 
where the coupling
between the external electromagnetic currents and the exchanged gluons
is mediated by three photon impact factors.

Let us start with the amplitude $T_{\gamma^*(pn) \to \gamma ^* (pn)}$.
The kinematics is illustrated in figure~\ref{fig:gammapn}: the
amplitude depends upon three energy variables, $s_1=(q+p_1)^2$,
$s_2=(q'+p'_2)^2$, and $M^2 = (q+p_1 - p'_1)^2$. In the high energy
limit that we are interested in, $s_1 \simeq s_2 \simeq s$ and all
these variables are of the same order as $S=(q+p_1+p_2)^2\simeq 2s$. All these
energies are assumed to be much larger than the momentum transfer
variables $t=(q-q')^2$, $t_1=(p_1-p'_1)^2$, and $t_2=(p_2-p'_2)^2$ and
the virtuality of the photon, $Q^2 = - q^2$,
\begin{equation}
  \label{eq:triple_Regge_kin}
  s_1, s_2 \gg M^2 \gg Q^2, -t_1, - t_2, -t \,.
\end{equation}
We will distinguish between $s_1$ 
and $s_2$, but at the end we set $s_1= s_2 =s$ and $t=0$. 
Throughout this paper we use Sudakov variables 
with the lightlike reference vectors $q_A$ and $p_B$, such that $s= 2q_A \cdot q_B$, 
$S= 4 q_A \cdot p_B =2s$, $q=q_A-xp_B$ with 
$x=Q^2/(2q_A\cdot p_B)$ and 
\begin{equation}
  \label{eq:M2}
    x_P = \frac{M^2 + Q^2 - t_1}{s} \simeq \frac{M^2}{s} \ll 1 \,.
\end{equation}
Neglecting the nucleon masses we have 
\begin{equation}
p_1 = p_2 =p_B\,,\:\:\:\:\:\:
p'_1 = p_B(1-x_P) +p_{1 \perp} \,,\:\:\:\:\:\:
p'_2= p_B(1+x_P) + p_{2 \perp} \,. 
\end{equation}
Internal momenta are then written as 
\begin{equation}
\label{eq:Sudakov}
k_i=\alpha_i q_A+ \beta_i p_B + k_{i\,\perp} 
\end{equation}
with $k^2_{i\,\perp} = - \bk_i^2$. The fact that the two nucleons are
in a weakly coupled bound state implies that we will allow the two
nucleons to have small losses of longitudinal and transverse momenta,
i.\,e.\ we will integrate over $x_P$ and $p_{1 \perp} = - p_{2 \perp}=
k_{\perp}$. The integration over $x_P$ is equivalent to the
integration over the mass squared $M^2$, and the latter will be kept
much smaller than $s$.

A convenient way of computing the elastic scattering amplitude
$T_{\gamma^*(pn) \to \gamma^* (pn)}$ in the high energy limit is to
use dispersion relations and Regge theory, for a review of these
techniques see \cite{Brower:1974yv}. For our case\footnote{This 
representation holds for the scattering of scalar particles. 
Because of helicity conservation (which is a consequence of 
supersymmetry), this representation can also be used for our case, 
where we consider the scattering of external vector currents.} the
scattering amplitude can be written in the form 
\be
\label{tripleregge}
\begin{split}
T_{3 \to 3} & (s_1, s_2, M^2; t_1,t_2,t) 
\\
& 
= \int \frac{dj_1 dj_2 dj}
{(2\pi i)^3} \, s_1^{j_1} \xi(j_1) {s_2}^{j_2} \xi(j_2) (M^2)^{j-j_1-j_2}
\xi(j,j_1,j_2)
F(j_1,j_2,j; t_1,t_2,t)  
\end{split}
\ee
with the signature factors $\xi(j) = - \pi \frac{e^{-i\pi j} +1}{\sin \pi j}$,
$\xi(j, j_1,j_2) = - \pi \frac{e^{-i\pi (j-j_1-j_2)} +1}{\sin \pi (j-j_1-j_2)}$.
Given the representation (\ref{tripleregge}), there is an easy way 
of computing this amplitude. Namely, we take the triple discontinuity 
in $s_1$, $s_2$, and $M^2$, 
\be
\label{tripledisc}
{\rm disc}_{s_1} \, {\rm disc}_{s_2} \, {\rm disc}_{M^2} T_{3 \to 3} =
\pi^3 \int \frac{dj_1 dj_2 dj}
{(2\pi i)^3} s_1^{j_1} {s_2}^{j_2} (M^2)^{j-j_1-j_2}
F(j_1,j_2,j; t_1,t_2,t) \,,  
\ee
and see that the partial wave $F(j_1,j_2,j; t_1,t_2,t)$ which 
in our kinematic region is real-valued (i.\,e.\ has no internal phases) can 
be computed from the triple Mellin transform of the (real-valued) triple 
energy discontinuity. Using unitarity, this triple energy discontinuity 
is easily obtained from high energy production processes. 

To obtain the cross section \eqref{eq:sigmatot} for deep inelastic 
scattering on the deuteron from eq.\ \eqref{tripleregge} it is needed to 
take the imaginary part, i.\,e.\ the discontinuity in $M^2$, 
to set $s_1=s_2 = s$ and $t=0$ (which implies 
$p_{1 \perp} = - p_{2 \perp}= k_{\perp}$), and to integrate over the 
phase space of the two nucleons, i.\,e.\ over $x_P =M^2/s$ and ${ k_\perp}$, 
\be
\label{triplereggeint}
\begin{split}
\sigma^{\text{tot}}_{\gamma^* (pn)\to }& {}_{\gamma^* (pn)}(s) = 
\frac{1}{2s}\int_x^1 {d x_P} \int\!\! \frac{d^2 {\bf k}}{2(2\pi)^3}
\, \text{disc}_{M^2} T_{3 \to 3}
(s, s,M^2 = x_Ps; -{\bf k}^2,-{\bf k}^2,0)
\\
=& {}\, \frac{1}{2}\int \frac{dj_1 dj_2 dj}
{(2\pi i)^3} \, s^{j-1} \xi(j_1) \xi(j_2) \,\frac{1}{j-j_1-j_2+1}
\int \!\!\frac{d^2 {\bf k}}{2(2\pi)^3} \, F(j_1,j_2,j; -{\bf k}^2, -{\bf k}^2, 0)
\,,
\end{split}
\ee
where due to the nucleon form factors the integration over ${\bf k}^2 =
-t_1 =-t_2$ remains restricted to a small range. 

For the discussion of this paper, however, we focus on the correlator
of six currents, $T_{3 \to 3}(s_1, s_2, M^2; t_1,t_2,t)$, and the
integrations over $x_P$ and ${\bf k}$ will not be considered. In the
following it will be convenient to introduce instead of the angular
momenta $j$, $j_1$, $j_2$ the variables $\omega=j-1$,
$\omega_1=j_1-1$, $\omega_2=j_2-1$, and we will write
$F(\omega_1,\omega_2,\omega; t_1,t_2,t)$ instead of $F(j_1,j_2,j;
t_1,t_2,t)$.

In order to form, in the leading logarithmic approximation, color singlet 
$t$-channel states which couple at the top 
to the virtual photon and at the bottom to two photons, one is 
led to QCD diagrams with four $t$-channel gluons at the lower end and 
two, three, or four gluons at the upper end. 
A few examples of such diagrams are shown in figure~\ref{fig:diagrams}. 
Wavy $t$-channel gluon lines stand for reggeized gluon 
propagators, and horizontal lines between the $t$-channel gluons denote 
on-shell $s$-channel gluons (that is real gluon production in the 
interaction kernel).%
\FIGURE{
\includegraphics[width=8.6cm]{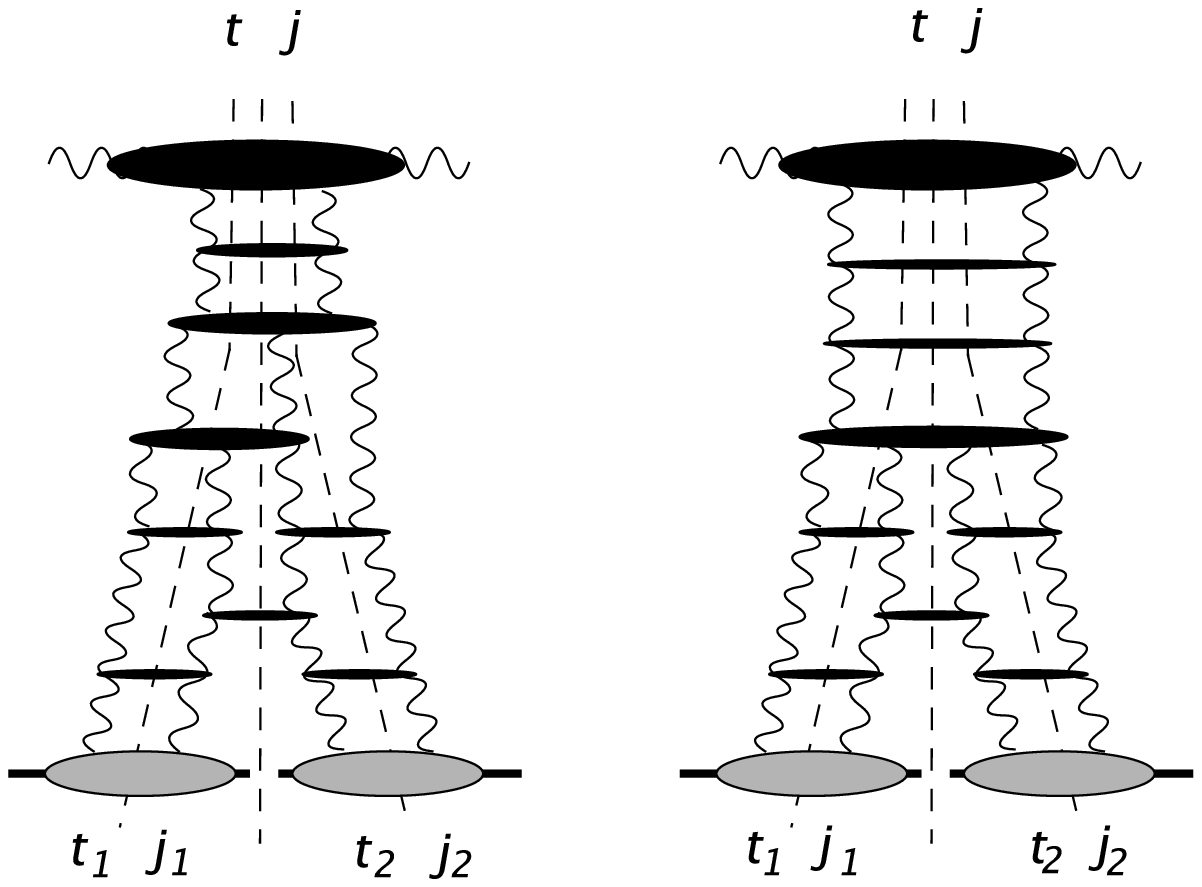}
\caption{A few contributions to the triple energy discontinuity in 
eq.\ (\ref{tripleregge})
\label{fig:diagrams}} 
} 
For the computation of the triple energy
discontinuity we proceed in the same way as for the LO BFKL
ladders. We use multiparticle amplitudes in the multi-Regge
kinematics: $T_{2\to n}$ and, more generally, $T_{n\to m}$, where all
incoming and outgoing particles are separated by large rapidity gaps.
These diagrams represent, for the six-point amplitude in the triple
Regge limit, the (generalized) leading logarithmic approximation: for
each gluon loop we have a logarithm of a large energy variable.

As to the general structure of the diagrams, at the lower end we start
from 4 reggeized gluons (two color singlets) which couple to the two
impact factors at the bottom. At the upper end, given by the upper
photon impact factor, we end with a $t$-channel state consisting of
two, three, or four gluons. We thus encounter $t$-channel states with
$2$, $3$, or $4$ gluons: their propagation is described by the BFKL
equation for the case of two gluons, and by the BKP equations in the
case of three and four gluons. When moving from the bottom to the
top, the number of $t$-channel gluons never increases. Transitions
between the different states are described by kernels $K_{2 \to 3}$
and $K_{2 \to 4}$ which we describe in section \ref{sec:eglla} below.
There are three different $t$-channels ($t$, $t_1$, $t_2$), and each
of them has its own angular momentum $\omega$, $\omega_1$, $\omega_2$,
respectively. As seen from figure~\ref{fig:diagrams}, there is always
a `lowest' interaction, which we call `branching vertex', below which
the diagrams split into the $t_1$ and $t_2$ channels. It is therefore
convenient to split the diagrams of figure~\ref{fig:diagrams} into
three pieces: below the branching vertex we have two disconnected BFKL
Pomerons, $D_{2}$, depending on $\omega_1$ and $\omega_2$,
respectively. At and above the vertex we have an amplitude with four
gluons, $D_{4}$, which depends upon $\omega$: it satisfies an integral
equation which, for the case of QCD, has been discussed in
\cite{Bartels:1994jj}, and has been further studied in
\cite{Bartels:1999aw}. One of the main results of that analysis is the
appearance of the M\"obius invariant $2 \to 4$ gluon vertex. Together
with the BFKL kernel it represents one of the fundamental building
blocks of QCD Reggeon field theory. The investigation of the
analogous amplitude in ${\cal N}=4$ SYM, $\Db_4(\omega)$, will be the
main goal of the present paper. In particular, we will study the
influence of the supersymmetric particle content of the impact factors
on the solution of the integral equation for that amplitude. 

It is straightforward to generalize this discussion of six-point amplitudes 
to eight-point, ten-point amplitudes etc. In the same way as the six-point 
amplitude leads to $4$ gluons in the $t$-channel, the eight-point amplitude 
contains up to $6$ gluons. As in the previous case, in QCD such amplitudes arise 
very naturally in the context of the scattering of a photon on nuclei 
consisting of three or more weakly bound nucleons. From the theoretical point 
of view, these multiparticle correlators provide a natural environment 
for color singlet BKP states. When deriving and analyzing these higher 
order BKP states within QCD, new transition vertices of reggeized gluons 
appear which are elements of QCD Reggeon field theory. In this paper we 
restrict ourselves to the six-point amplitude containing four gluons. We will, 
however, present a few results also on the five- and six-gluon states. 

\boldmath
\subsection{Six-point correlators of $R$-currents in ${\cal N}=4$ SYM}
\unboldmath

After this brief review of QCD calculations we now want to turn to
analogous scattering amplitudes in ${\cal N}=4$ SYM. In terms of
component fields, this theory contains the vector field $A_\mu$, 4
chiral spinors $\lambda_I$, and 6 real scalars $X_M$. They all
transform in the adjoint representation of the gauge group
$\text{SU}(N_c)$, and generically we can write the fields as
$\Phi=\Phi_{ab}=\Phi^c (T^c)_{ab}$, with the generators of the adjoint
representation, $(T^c)_{ab}=-i f_{abc}$. The $f_{abc}$ are the
$\mbox{SU}(N_c)$ structure constants which occur also in the algebra
of the generators $t^a$ of the fundamental representation,
$[t^a,t^b]=i f_{abc} t^c$. Our convention for the normalization of
the $t^a$ is such that $\tr(t^at^b)=\delta_{ab}/2$. For the generators
in the adjoint representation we have $[T^a,T^b]=i f_{abc} T^c$ and
$\tr (T^a T^b)= N_c \delta_{ab}$.

The Lagrangian of ${\cal N}=4$ SYM theory is \cite{Brink:1976bc} 
\begin{equation}
  \label{eq:LSYM4}
  \begin{split}
    \cL = \tr\Big(&
    - \frac{1}{2} F_{\mu\nu}F^{\mu\nu}
    +  D_\mu X_M D^\mu X_M
    + 2 i \lambda_I \sigma^\mu D_\mu \lambdab^I \\
    &-2 i g \lambda_I [\lambda_J,X^{IJ}]
    - 2 i g \lambdab^I [\lambdab^J,X_{IJ}]
    + \frac{1}{2} g^2 [X_M,X_N][X_M,X_N]
    \Big) \,. 
  \end{split}
\end{equation}
The $X_M$ and $X_{IJ}$ are related by the $\text{SU}(4)\cong
\text{SO}(6)$ sigma symbols,
\begin{equation}
  \label{eq:SigmaSymb}
  X_{IJ} = -\frac{1}{2} (\Sigma_M)_{IJ} X_M \, , \qquad\qquad
  X^{IJ} = \frac{1}{2} (\Sigma^{-1}_M)^{IJ} X_M \,,
\end{equation}
with $\tr(\Sigma_M\Sigma_N^{-1}) = 4 \delta_{MN}$, which implies that
$X_M X_M = X_{IJ} X^{IJ}$.
Capital indices transform under the $R$-symmetry group $\text{SU}(4)$. 
In particular,
$A,B,C,...=1,...,15$ are indices of the adjoint representation,
$I,J,K,... =1,...,4$ transform under the fundamental, and
$M,N,...=1,...,6$ under the vector representations of the $R$-symmetry.
Small indices $a,b,c,...=1,...,N_c^2-1$ are adjoint representation
indices for the gauge group $\text{SU}(N_c)$.
The covariant derivative $D_\mu$ and the gauge field strength tensor
$F_{\mu\nu}$ are defined in the usual way by (writing $\Phi$ generically 
for any field in the theory)
\begin{eqnarray}
  \label{eq:DmuFmunu}
  D_\mu \Phi &=&{} \partial_\mu \Phi - i g [A_\mu, \Phi] \, ,\\
  F_{\mu\nu} &=&{} \partial_\mu A_\nu - \partial_\nu A_\mu
                 - i g [A_\mu, A_\nu] \,.
\end{eqnarray}

The theory enjoys a $\mbox{SU}_R(4)$ global symmetry, called
$R$-symmetry, which transforms the different supercharges. Under these
transformations, the fields $A_\mu$, $\lambda_I$, $X_M$ belong to the
scalar, fundamental, and vector representation, respectively. More
specifically, the Lagrangian \eqref{eq:LSYM4} is invariant under
the global $R$-symmetry transformation
\begin{equation}
  \label{eq:Rsymtransf}
\begin{array}{lcl}
    \delta \lambda^{a \alpha I} & = &{} i \epsilon_A
    \lambda^{a \alpha J} (T^A)_{JI} \,,\\
    \delta \lambdab^{a \dot{\alpha} I} & = &{} - i \epsilon_A
    (T^A)_{IJ} \lambdab^{a \dot{\alpha} J} \,,\\
    \delta X^a_M & = &{}  i \epsilon_A (T^A)_{M N} X^a_N \,,
  \end{array}
\end{equation}
where $\epsilon_A$ are small parameters, and $T^A$ are the $\mbox{SU}_R(4)$ 
generators in the appropriate representation.\footnote{In our notation the 
generators of $\mbox{SU}_R(4)$ are labeled by capital letters so that they 
are distinguished from the generators of $\mbox{SU}(N_c)$ which carry 
small letters.} 
The corresponding Noether current is 
\begin{equation}
  \label{eq:Rcurrent}
  J_R^{A \mu} =
  i \frac{\partial \cL}{\partial(\partial_\mu \Phi)} \Delta^A \Phi =
  \tr \left(
  - \lambda \sigma^\mu T^A \lambdab
  - i  X T^A D^\mu X \right) \,,
\end{equation}
where $\Delta^A \Phi$ is obtained from \eqref{eq:Rsymtransf} with
the definition $\delta\Phi = i \epsilon_A \Delta^A \Phi$ for an
infinitesimal $R$-transformation.

In \cite{CaronHuot:2006te} it has been suggested that in ${\cal N}=4$
SYM this global current can be used as a substitute for the
electromagnetic $\mbox{U}(1)$ current in QCD. To be more precise, one
should choose an abelian subgroup of the global $\text{SU}_R(4)$, for
example the one generated by $T^3 = \mbox{diag}
(\frac{1}{2},-\frac{1}{2},0,0)$. The simplest application of this
procedure is the supersymmetric analog of elastic $\gamma^* \gamma^*$
scattering: the elastic scattering of two $R$-currents
\cite{Bartels:2008zy}. In QCD this process, when evaluated at energies
much larger than the photon virtualities, provides one of the cleanest
environments for studying the BFKL Pomeron. With the conjectured
AdS/CFT duality, the correlator of four $R$-currents therefore offers
the possibility to study the dual of the BFKL Pomeron on the string
theory side.

As a next step along this line, one may address higher correlators,
e.\,g.\ the six-point function for which the QCD side has been
discussed above. Giving labels $A$, $B_1$, $B_2$ to the three
incoming virtual photons (and analogous primed labels to the outgoing
ones) in the amplitude $T_{\gamma^* \gamma^* \gamma^*\to \gamma ^*
  \gamma^* \gamma^*}$, one is led to consider an analogous process in
${\cal N}=4$ SYM by defining the momentum space six-point function
(see figure~\ref{fig:3r_current})
\begin{align}
  \label{eq:6pointfunc}
    i (2\pi)^4 &\delta(q+ p_1+p_2-q'-p'_1-p'_2) \,
T_{3 \to 3}^{\mu_A \mu_{B_1} \mu_{B_2}\mu_{A'} \mu_{B'_1}\mu_{B'_2}}
\notag \\
=& \prod_{i=A,...,B'_2} \int d^4x_i \,
    e^{- i q \cdot x_A - i p_1 \cdot x_{B_1} - i p_2 \cdot x_{B_2}
      +i q' \cdot x_{A'} + i p'_1 \cdot x_{B'_1} + i p'_2 \cdot x_{B'_2}}
\notag \\
&
\langle 
J_R^{A\mu_A}(x_A) J_R^{B_1\mu _{B_1}}(x_{B_1}) J_R^{B_2\mu _{B_2}}(x_{B_2})   
J_R^{A'\mu_{A'}}(x_{A'}) J_R^{B'_1\mu_{B'_1}}(x_{B'_1}) J_R^{B'_2\mu_{B'_2}}(x_{B'_2}) 
\rangle 
\,.
\end{align}
\FIGURE{
  \includegraphics[width=7cm]{gggto3n4.epsi}
  \caption{The momentum space six-point function of $3 \to 3$ $R$-current scattering. 
  \label{fig:3r_current}}
}
Following the discussion of the previous subsection, we will be 
interested in the (generalized) leading logarithmic approximation of
the six-point function $T_{3 \to 3 } $ in the triple
Regge limit, where we will make use of the analytic representation \eqref{tripleregge}.

The only places where the supersymmetric content of ${\cal N}=4$ SYM
becomes visible in the above expression are the impact factors.
Compared to the QCD case, there are two novel features:(i) the (Weyl)
fermions are in the adjoint representation, (ii) in addition to the
fermion loop, we have the scalars which occur in the adjoint
representation as well. In \cite{Bartels:2008zy} these impact factors
have been calculated for the four-point function where two $t$-channel
gluons are coupled to the external currents. For the six-point
function, new impact factors with three or four $t$-channel gluons
appear as well. They have not been calculated yet, and their
computation constitutes one of the main goals of this paper.

The structure of these novel impact factors has quite important
consequences. In the QCD analysis the integral equations which
formally sum all the Reggeon diagrams can partially be solved and
simplified. The key ingredients to this are the reggeization of the
gluon and the validity of bootstrap equations. The latter ones
strongly depend upon the structure of the impact factors which -- in
QCD with quarks in the fundamental representation -- are simple
(Dirac) fermion loops. In QCD it has been shown that all
contributions with more than two $t$-channel gluons are absorbed into
reggeizing pieces, and, at the end, only two-gluon contributions
remain. This result is closely connected with the structure of the $2
\to 4$ gluon vertex. In the case of ${\cal N}=4$ SYM the fermions are
in the adjoint representation, and the impact factors also contain
contributions from the scalars: the structure of the impact factors
with three or more $t$-channel vector particles is different, and it
is {\sl a priori} not obvious how this affects the solution of the
integral equations. We shall investigate this issue in the present
paper.

\section{Impact factors}
\label{sec:external}

\boldmath
\subsection{Four-point functions in ${\cal N}=4$ SYM}
\unboldmath

It will be useful to briefly recapitulate the two-gluon impact factor
which was studied in detail in \cite{Bartels:2008zy}, some properties
of this impact factor were also discussed in
\cite{Cornalba:2008qf,Cornalba:2009ax}. Its precise definition is
given by 
\be
\label{eq:def_impact2}
\Db_{(2;0)}^{a_1a_2,\lambda_A\lambda_{A'}} (\kf_1, \kf_2) 
= \,
\frac{p_B^{\rho_1}p_B^{\rho_2}}{s^2} \int_{0}^\infty \frac{d\tilde{s}}{2\pi} \,
\text{disc}_{\tilde s} \,
A^{\mu_A \mu_{A'},\rho_1 \rho_2;a_1a_2}_{R_Ag_1 \to R_{A'}g_2 }
\epsilon^{\lambda_{A}}_{\mu_{A}}(q)\epsilon^{\lambda_{A'}}_{\mu_{A'}}(q') 
\,.
\ee
Here $A^{\mu_A \mu_{A'},\rho_1 \rho_2;a_1a_2}_{R_Ag_1 
\to R_{A'}g_2 } $ is the amplitude for scattering of the
$R$-current $A$ with Lorentz index $\mu_A$ and a gluon with momentum
$-k_1$, Lorentz index $\rho_1$ and color label $a_1$ into the
$R$-current $A'$ with Lorentz index $\mu_{A'}$ and a gluon with momentum
$k_2$, Lorentz index $\rho_2$ and color label $a_2$. $\tilde s = (q -
k_1)^2 \simeq q^2 -{\bf k}_1^2 - s \beta$ is the total center-of-mass
energy squared of the $R$-current-gluon system.
$\epsilon_\mu^\lambda(k)$ are the polarization vectors of the
$R$-currents with polarizations $\lambda = L, h$ where $L$ denotes
longitudinal and $h=1,2$ transverse polarizations.

We begin our discussion with the fermionic part of the $R$-current
impact factor which consists of Weyl fermions in the adjoint
representation of $\text{SU}(N_c)$. Compared to Dirac fermions in the
fundamental representation (that is the usual QCD case), we have to
consider the following changes. Instead of $N_c$ fundamental quarks we
now have $N_c^2-1$ adjoint particles, i.\,e.\ the color trace
$\tr(t^at^b)=\delta_{ab} /2 $ is replaced by $\tr(T^aT^b)=N_c
\delta_{ab}$ where $(T^a)_{bc} = -i f_{abc}$ are the generators in the
adjoint representation. Next, we have to consider the $\mbox{U}(1)$
charges $e_F$ of the global $\text{SU}_R(4)$ symmetry of the Weyl
fermions which are the analogs of the electric charges $e_q$ in QCD.
With our choice of the $\text{U}(1)$ subgroup, $T^3 = {\rm diag}
(\frac{1}{2},-\frac{1}{2},0,0)$, we can take care of these charge
factors by multiplying the QCD amplitude by
\begin{equation} 
\label{RFdef}
R_F= \frac{\sum e_{F}^2}{\sum e_q^2} =\frac{1}{2 \sum e_q^2} \,, 
\end{equation}
since here $\sum e_{F}^2= \tr_4 (T^A T^A) = \frac{1}{2}$. 
Furthermore, we identify the left- and right-handed components of a massless 
Dirac fermion with Weyl fermions in the standard way, and we can conclude 
that the impact factor with a massless Dirac fermion is twice the 
corresponding impact factor with a Weyl fermion. 
Compared to a Dirac fermion in the fundamental representation in QCD, we therefore 
have for the fermionic contribution in ${\cal N}=4$ SYM the relative weight
\be
\label{2NcR}
2 N_c R \quad \quad \mbox{with} \quad \quad R=\frac{1}{2} R_F =
\frac{1}{4 \sum e_q^2} \,. 
\ee 
The momentum structure, including the
integration over the loop momentum, on the other hand, remains the
same in each individual diagram and is not affected by the change of
the color representation of the quarks. Finally, we have the scalar
contribution for which there is no counterpart in QCD.

In the following we often compare with the QCD case, i.\,e.\ with the
case of Dirac fermions in the fundamental representation. In order to
make a clear distinction between the impact factors (and, later on,
also the gluon amplitudes) in the two theories we will denote
$n$-gluon impact factors in QCD with fundamental quarks by normal
letters, for example $D_{(n;0)}$, while those in ${\cal N}=4$ SYM will
be denoted by blackboard-style letters, for example $\Db_{(n;0)}$.

The full ${\cal N}=4$ SYM impact factor 
$\Db_{(2;0)}({\bf k}_1,{\bf k}_2)$ for the scattering of two $R$-currents 
with two exchanged
gluons in the $t$-channel has been computed in \cite{Bartels:2008zy}.
The computation of the fermionic part includes the four diagrams shown
in figure~\ref{fig:IFFermions}.%
\FIGURE{
 \includegraphics[width=12cm]{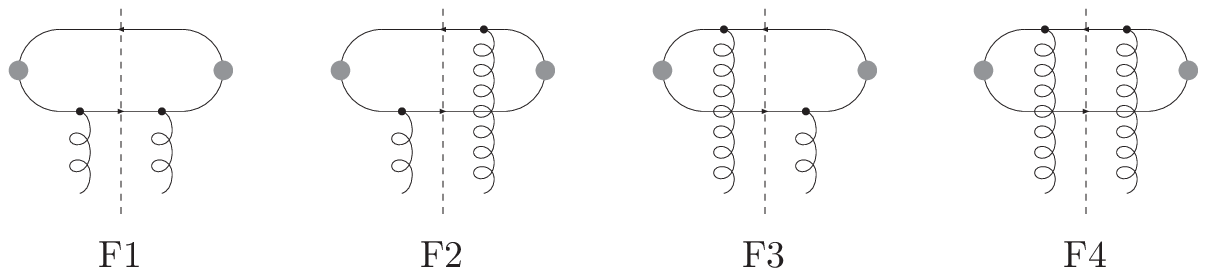}
  \caption{The fermion diagrams for the impact factors.
    \label{fig:IFFermions}} 
} 
We consider the discontinuity in $s$ and
therefore two propagators are set on-shell. Due to the cut only
diagrams in which the gluon lines do not cross have to be
included. The external currents are projected onto different
polarizations, longitudinal or transverse. For simplicity we give the
result of the fermion impact factor, the sum of the four diagrams in
figure~\ref{fig:IFFermions}, only for transverse polarization of the
$R$-currents:
\begin{eqnarray}
\Db_{F,(2;0)}^{a_1a_2,hh'}({\bf k}_1,{\bf k}_2)&=&{}
\delta_{a_1a_2}\,\frac{N_c \alpha_s}{2} 
 \int_0^1 d\alpha\int\frac{d^2{\bl}}{(2\pi)^2}\nonumber\\
&&{}
\times\left[-4 \alpha(1-\alpha) {\beps}^{(h)}\cdot\left(\frac{\bN_1}{D_1}
-\frac{\bN_2}{D_2}\right)\left(\frac{\bN_1'}{D_1'}-\frac{\bN_2'}{D_2'}\right)
\cdot{\beps}^{(h')*}\right.\nonumber\\
&&{} \hspace*{0.7cm}
+\left.{\beps}^{(h)}\cdot{\beps}^{(h')*}\left(\frac{\bN_1}{D_1}
-\frac{\bN_2}{D_2}\right)\left(\frac{\bN_1'}{D_1'}
-\frac{\bN_2'}{D_2'}\right)\right] \,.
\end{eqnarray}
Here $h$ and $h'$ denote the transverse polarizations (which we will
often suppress in the notation of the amplitudes $\Db_n$) and
$\beps^{(h)}$ denotes the corresponding polarization vector. $a$ and
$a'$ are the color labels, and ${\bf k}_1$ and ${\bf k}_2$ with ${\bf
  q} = {\bf k}_1 + {\bf k}_2$ are the transverse momenta of the
gluons. The trace over the two generators of the $\text{SU}_R(4)$
group is included in the impact factor. For fermions in the
fundamental representation it is
\begin{equation}
\tr_4(T^AT^A)=\frac{1}{2} \,,
\end{equation}
as it appeared already in the relative factor $R_F$, see
(\ref{RFdef}). The integrations which are left are over the
transverse momentum $\bl$ and the Sudakov component $\alpha$,
belonging to the longitudinal momentum $q_A$, see (\ref{eq:Sudakov}),
of the fermion loop. The propagators and numerators are given by ($i
\in \{ 1,2 \}$) \be
\label{eq:Prop}
\begin{split}
 \bN_{1} &= \bl \\
\bN'_{1} &= \bl-(1-\alpha)\bq \\
\bN_{2} &= \bl-\bk \\
\bN'_{2} &= \bl-\bk+\alpha\bq \\
D_i &= \bN_i^2 + \alpha (1-\alpha) Q_A^2 \\
D'_i &= \bN'^2_i + \alpha (1-\alpha) Q_{A'}^2 \,.
\end{split}
\ee 
Note that $\Db_{F,(2;0)} (\kf_1, \kf_2)$ is symmetric in its two
momentum arguments and vanishes if one of them vanishes ($i \in
\{1,2\}$): 
\be
\label{Dbf20nullstell}
\left. \Db_{F,(2;0)}^{a_1a_2} (\kf_1, \kf_2)\right|_{\kf_i=0} = 0
\,, 
\ee
which is a consequence of the gauge invariance of the impact factor. 

The scalar contribution to the impact factor in ${\cal N}=4$ SYM
consists of nine diagrams, shown in figure~\ref{fig:IFScalars}, and
all diagrams are necessary to satisfy the Ward identities at finite
energies. But at high energies the diagrams S5-S9 are suppressed
\cite{Bartels:2008zy}, and the leading diagrams for the scalar impact
factor are S1-S4, which are similar to the fermionic ones.%
\FIGURE{
   \includegraphics[width=10cm]{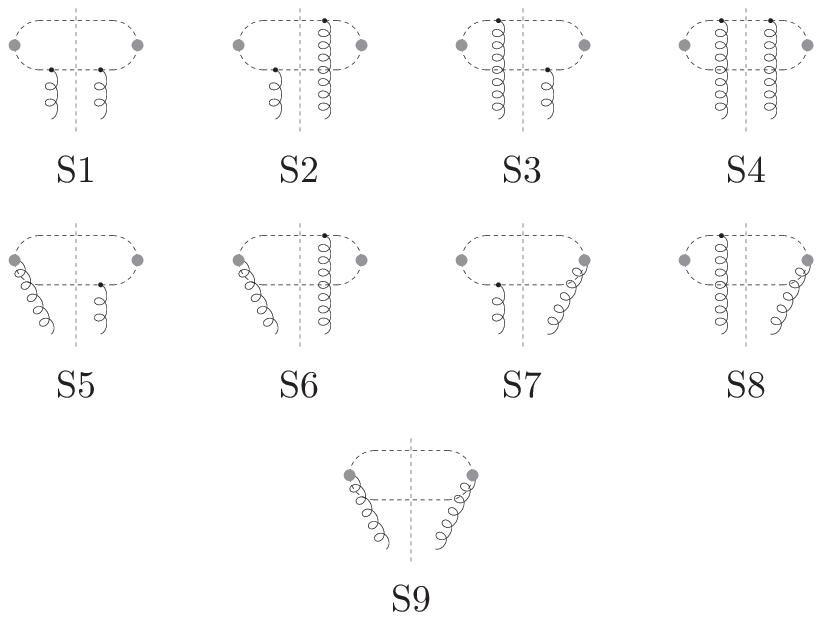}
  \caption{The scalar diagrams for the impact factors.
    \label{fig:IFScalars}}
}
The scalar part of the impact factor with transversely polarized $R$-currents is
\begin{eqnarray}
\Db_{S,(2;0)}^{a_1a_2,hh'}({\bf k_1},{\bf k_2})&=&{}
2\delta_{a_1a_2}\,N_c\alpha_s \int_0^1 d\alpha\int\frac{d^2{\bl}}{(2\pi)^2} \,\alpha(1-\alpha)
\nonumber\\
&&{}
\times \,{\beps}^{(h)}\cdot\left(\frac{\bN_1}{D_1}
-\frac{\bN_2}{D_2}\right)\left(\frac{\bN_1'}{D_1'}-\frac{\bN_2'}{D_2'}\right)\cdot{\beps}^{(h')*}
\end{eqnarray}
with the propagators and numerators given in (\ref{eq:Prop}). The
trace over the $\text{SU}_R(4)$ generators for scalars in the vector
representation, that is included here, gives
\begin{equation}
\tr_6(T^AT^A)=1 \,,
\end{equation}
different from the fermionic case. Also the scalar contribution
$\Db_{S,(2;0)} (\kf_1, \kf_2)$ to the impact factor is symmetric in
its two momentum arguments and vanishes if one of them vanishes ($i
\in \{1,2\}$): \be
\label{Dbs20nullstell}
\left. \Db_{S,(2;0)}^{a_1a_2, hh'} (\kf_1, \kf_2)\right|_{\kf_i=0} = 0
\,. 
\ee

We obtain the full impact factor in ${\cal N}=4$ SYM as 
\begin{equation}
\Db^{a_1a_2,hh'}_{(2;0)}\equiv \Db^{a_1a_2,hh'}_{F,(2;0)}+\Db^{a_1a_2,hh'}_{S,(2;0)} \,,
\end{equation}
and obtain for the example of transversely polarized $R$-currents 
\be
\label{eq:fullimpacttwogluon}
\Db^{a_1a_2,hh'}_{(2;0)}=\delta_{a_1a_2}\delta_{hh'}\frac{N_c\alpha_s}{2}
\int_0^1d\alpha\int\frac{d^2\bl}{(2\pi)^2}\left(\frac{\bN_1}{D_1}-
\frac{\bN_2}{D_2}\right)\cdot\left(\frac{\bN_1'}{D_1'}-\frac{\bN_2'}{D_2'}\right) \,.
\ee
$\Db_{(2;0)}$ is symmetric in its two momentum arguments, and 
as a consequence of (\ref{Dbf20nullstell}) and (\ref{Dbs20nullstell}) we have 
\be
\label{Db20nullstell}
\left. \Db_{(2;0)}^{a_1a_2,hh'} (\kf_1, \kf_2)\right|_{\kf_i=0} = 0
\,. 
\ee 
 
It has been observed in \cite{Bartels:2008zy} that due to supersymmetry
the helicities of the scattering $R$-currents are conserved. In particular,
unlike the QCD case, helicity conservation holds for the $R$-current
impact factor also in the non-forward direction $t \neq 0$.
In the following we will keep in our notation the different
polarizations $\lambda_A$ and $\lambda_{A'}$ explicit, while we keep in mind that the
impact factor is always proportional to 
$\delta^{\lambda_{A}\lambda_{A'} }$. This also justifies a posteriori
the use of the analytic representation \eqref{tripleregge} for
the analysis of the six point function (\ref{eq:6pointfunc}).

\boldmath
\subsection{Six-point functions in ${\cal N}=4$ SYM}
\label{sixpointinn4sym}
\unboldmath

The next step is to go to higher correlation functions, e.\,g.\
six-point functions (\ref{eq:6pointfunc}). Again fermions and
scalars contribute to the impact factor $\Db_{(n;0)}$. They
generalize \eqref{eq:def_impact2} to an arbitrary number of
gluons and are defined as
\be
\begin{split}
  \label{eq:def_impactn}
  \Db&{}_{(n;0)}^{a_1a_2\ldots a_n,\lambda_A\lambda_{A'}} (\kf_1,\ldots, \kf_n) 
\\
&=
\frac{p_B^{\rho_1} \ldots p_B^{\rho_n}}{s^{n}} 
\int_{0}^\infty \frac{d\tilde{s}_1}{2\pi}\ldots  
 \int_{0}^\infty \frac{d\tilde{s}_{n-1}}{2\pi}\,
 \text{disc}_{\tilde {s}_1}\ldots  \text{disc}_{\tilde {s}_{n-1}}
 A^{\mu_A \mu_{A'},\rho_1 \ldots \rho_n;a_1\ldots a_n}_{R_Ag_1 \to R_{A'}g_2\ldots g_n } 
\epsilon^{\lambda_{A}}_{\mu_{A}}(q)\epsilon^{\lambda_{A'}}_{\mu_{A'}}(q') \,,
\end{split}
\ee
with $\tilde{s}_i = (q - \sum_{j=1}^{i}k_j)^2$, $i = 1, \ldots n-1$. One
possible diagram with fermion loops contributing to the six-point
function is depicted in figure~\ref{fig:Fermion6}. The complete
impact factors are again given by the sum over all possible ways in
which the gluons can couple to the fermion and scalar
lines.% 
\FIGURE{ \hspace{4cm}
\includegraphics[width=5cm]{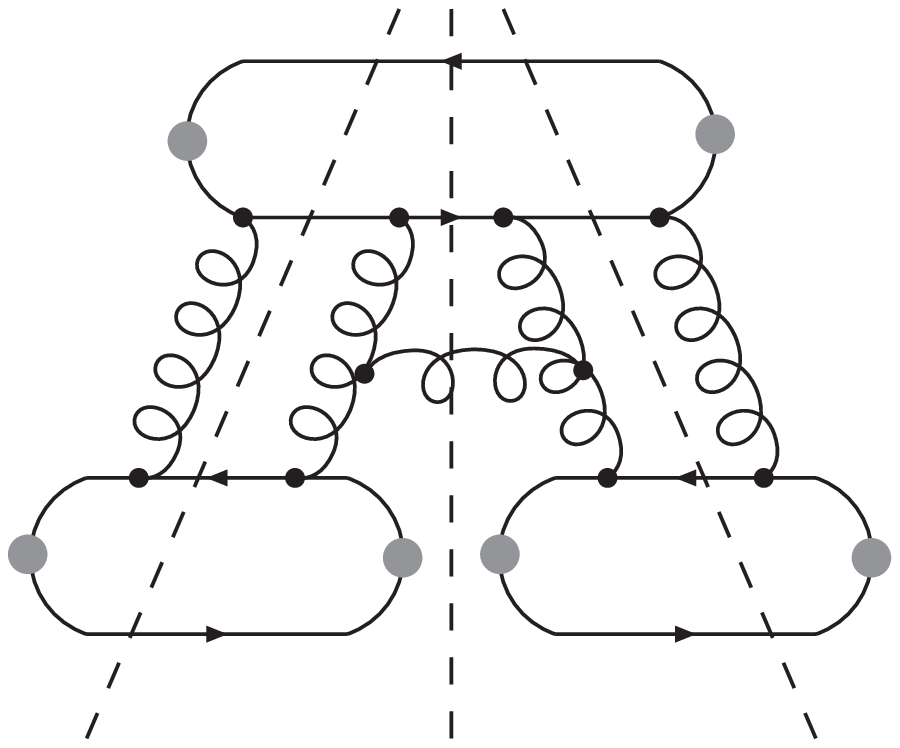}
\hspace{4cm}
\caption{A fermionic contribution to the six-point function
\label{fig:Fermion6}}
}

In the following discussion of amplitudes with more gluons the color
factors will play a crucial role, in particular when we explain how
amplitudes with different numbers of gluons are related to each other
via reggeization. We will invoke known results from QCD in order to
derive these relations for the case of ${\cal N}=4$ SYM. When
comparing the fermionic contributions we will always find the overall
factor $R$ of (\ref{2NcR}) when comparing an ${\cal N}=4$ SYM
amplitude to its analog in QCD. The color factor, on the other hand,
will have a richer structure, such that the main structural difference
between the two theories originates from the color representations of
the particles. In the following we will therefore sometimes (in
slight abuse of language) speak of the `adjoint' and `fundamental'
representation when we actually refer to ${\cal N}=4$ SYM and QCD,
respectively. In fact, the results obtained below for the fermions
can be used for considering a theory like QCD with adjoint instead of
fundamental quarks by just dropping the factor $R$ wherever it occurs.

\subsubsection{Fermionic impact factor}
\label{sec:fermionimpact}

For the case of fundamental quarks in QCD the impact factors with up
to six gluons have been given explicitly in
\cite{Bartels:1994jj,Bartels:1999aw}. There it has been found that
the impact factors with arbitrarily many gluons can be related to the
two-gluon impact factor $D_{(2;0)}$. A detailed account of how this
reggeization of the impact factors results from the corresponding
diagrams has been given in \cite{Braunewell:2005ct}. Here we want to
relate the Weyl fermion impact factors in the adjoint representation
to those of the fundamental representation. In accordance with the
notation of the previous section we assign color labels $a_i$ and
transverse momenta $\kf_i$ to the gluons.

To understand the main difference between QCD and ${\cal N}=4$ SYM we
have to take a closer look at the traces in color space. Inspection of
the possible diagrams contributing to the $n$-gluon impact factor
shows that these diagrams come in pairs: for each diagram there is
another diagram with the same momentum integration but with the
generators occurring in opposite order in the trace in color
space. The relative sign between these two diagrams turns out to be
positive for even numbers of gluons and negative for odd numbers of
gluons. It further turns out that the full impact factors can be
written completely in terms of the momentum part of the two-gluon
impact factor, see \cite{Bartels:1994jj,Bartels:1999aw}. (In that
representation the diagrams with all gluons attached to the same quark
line occur several times with different color factors but cancel in
the sum over two-gluon impact factors such that the original two
diagrams of this type are correctly counted.) It is straightforward to
check, following for example the derivation presented in
\cite{Braunewell:2005ct}, that this result holds also in the case of
adjoint fermions.

Let us first define the momentum part of the two-gluon amplitude by
separating it from the color tensor, \be
\label{sepcolmom}
D_{(2;0)}^{a_1 a_2} (\kf_1,\kf_2) = 
\delta_{a_1 a_2} D_{(2;0)}  (\kf_1,\kf_2) \,,
\ee
and analogously for $\Db_{(2;0)}^{a_1 a_2}$. 

For three gluons a pair of diagrams as described above comes with the
difference of two traces over generators of the respective
representation. For the adjoint representation we have for instance
\be
\label{three_colour:adj}
\tr(T^{a_1}T^{a_2}T^{a_3}) - \tr(T^{a_3} T^{a_2} T^{a_1}) 
= i N_c f_{a_1a_2a_3}\,,
\ee
while in the fundamental representation (i.\,e.\ in QCD) we had 
\be
\label{three_colour:fund}
\tr(t^{a_1}t^{a_2}t^{a_3}) - \tr(t^{a_3} t^{a_2} t^{a_1}) =
\frac{i}{2} f_{a_1a_2a_3} 
\ee 
instead, giving again rise to a relative
factor $2 N_cR$ in the ${\cal N}=4$ SYM case as compared to the QCD
case. This applies to all pairs of diagrams. Invoking the known
decomposition of the fundamental impact factor $D_{(3;0)}$ into a sum
over $D_{(2;0)}$ \cite{Bartels:1993ih,Bartels:1994jj}, \be
\label{D30_fund}
D_{(3;0)}^{a_1a_2a_3}(\kf_1,\kf_2,\kf_3) 
= \frac{1}{2} g f_{a_1a_2a_3} \,
[ D_{(2;0)}(12,3) - D_{(2;0)}(13,2) + D_{(2;0)}(1,23) ]  
\,,
\ee
we find for the adjoint representation in ${\cal N}=4$ SYM 
\be
  \label{D30_adj}
\begin{split}  
\Db_{F,(3;0)}^{a_1a_2a_3}(\kf_1,\kf_2,\kf_3) 
&=  2 N_c R \,D_{(3;0)}^{a_1a_2a_3}(\kf_1,\kf_2,\kf_3) 
\\
&= \fr{1}{2} g f_{a_1a_2a_3} \,
[ \Db_{F,(2;0)}(12,3) - \Db_{F,(2;0)}(13,2) + \Db_{F,(2;0)}(1,23) ]  \,. 
\end{split}
\ee Here we have made use of the shorthand notation for the momentum
arguments of $\Db_{(2;0)}$ and $D_{(2;0)}$ originally introduced in
\cite{Bartels:1999aw} in which the momenta are replaced by their
indices, and a string of indices stands for the sum of momenta, for
example 
\be
\label{notation}
D_{(2;0)} (12, 3) = D_{(2;0)}(\kf_1 + \kf_2, \kf_3) \,. 
 \ee
 In the
following this notation will also be used for other functions. Note
that we have expressed both three-gluon impact factors here in terms
of the momentum part of the two-gluon impact factor which does no
longer contain the color factor $\delta_{ab}$. We observe that the
three-gluon impact factor in the adjoint representation
\eqref{D30_adj} differs from the one in the fundamental
representation \eqref{D30_fund} only by a factor $2 N_c R$. As a
consequence the relation between the three-gluon impact factor and the
corresponding two-gluon impact factor is the same in both
representations, compare the second line of (\ref{D30_adj}) with
(\ref{D30_fund}). What we observe here is the reggeization of the
gluon in the impact factor. In each term in the sum in 
(\ref{D30_adj}) or (\ref{D30_fund}) two gluons combine to act
as a single gluon.

For four gluons the situation becomes more interesting. In the
fundamental representation, the color structure of a typical pair of
diagrams with the same momentum structure is given by the tensor 
\be
\label{d_abcd}
d^{abcd} = \tr(t^a t^b t^c t^d) + \tr(t^d t^c t^b t^a) 
\,.
\ee
Taking the quarks to be in the adjoint representation gives us the same 
two traces with the fundamental generators $t^a$ replaced by 
adjoint generators $T^a$, 
\be
\label{d_abcd_adj}
\tr(T^a T^b T^c T^d) + \tr(T^d T^c T^b T^a) 
= 2N_c d^{abcd} + 
\delta_{ab} \delta_{cd} + \delta_{ac}\delta_{bd} + \delta_{ad}\delta_{bc} 
\,,
\ee
where we have used \eqref{contr4f} and \eqref{zweimaltrace} 
in order to express 
this sum of traces in terms of the tensor in (\ref{d_abcd}) known 
from the fundamental representation. We notice that in the case 
of four gluons we find a further part in addition to reproducing 
$2 N_c$ times the tensor from the fundamental representation. 
Using the known result for the four-gluon impact factor in the fundamental 
representation \cite{Bartels:1994jj}, 
\bea
\label{d40}
D_{(4;0)}^{a_1a_2a_3a_4}(\kf_1,\kf_2,\kf_3,\kf_4)
&=&{} 
- g^2 d^{a_1a_2a_3a_4} \, [ D_{(2;0)}(123,4) + D_{(2;0)}(1,234)
                                  - D_{(2;0)}(14,23) ]
\nn \\
& &{} - g^2 d^{a_2a_1a_3a_4}  \, [ D_{(2;0)}(134,2) + D_{(2;0)}(124,3)
                                 - D_{(2;0)}(12,34) \nn \\
&&{} \hspace{2.4cm} - D_{(2;0)}(13,24) ]
\,,
\eea
and (\ref{d_abcd_adj}) we hence arrive at 
\bea
\label{D40_adj}
\lefteqn{ \Db_{F,(4;0)}^{a_1a_2a_3a_4}(\kf_1,\kf_2,\kf_3,\kf_4)
 \notag } \\
&=& {}   2 N_c R \, D_{(4;0)}^{a_1a_2a_3a_4}(\kf_1,\kf_2,\kf_3,\kf_4) +
\Db^{ a_1a_2a_3a_4}_{F,(4;0)\,\text{dir}} (\kf_1, \kf_2, \kf_3, \kf_4)
\nn \\
&=& {}   - g^2 d^{a_1a_2a_3a_4} \, [ \Db_{F,(2;0)}(123,4) +
\Db_{F,(2;0)}(1,234) - \Db_{F,(2;0)}(14,23) ]
\nn \\
& & {}   - g^2 d^{a_2a_1a_3a_4} \, [ \Db_{F,(2;0)}(134,2) +
\Db_{F,(2;0)}(124,3) - \Db_{F,(2;0)}(12,34) - \Db_{F,(2;0)}(13,24) ] 
\nn \\
& & {}   + \Db^{ a_1a_2a_3a_4}_{F,(4;0) \,\text{dir}} (1,2,3,4) \,.
\eea 
Here the additional part $\Db_{F,(4;0) \,\text{dir}}$ originates from the 
additional delta tensors in (\ref{d_abcd_adj}): later on it will be shown 
that this piece -- in contrast to the other terms in eq.\ \eqref{D40_adj} -- 
gives rise to a direct coupling of the four-gluon state in the $t$-channel 
to the external currents. Explicitly, it becomes 
\be
\label{D40_add}
\begin{split}
\Db^{a_1a_2a_3a_4}_{F,(4;0)\,\text{dir}} & (\kf_1, \kf_2, \kf_3, \kf_4)
\\
=&
-g^2 \frac{1}{2 N_c}\, 
(\delta_{a_1a_2} \delta_{a_3a_4} 
+ \delta_{a_1a_3}\delta_{a_2a_4} + \delta_{a_1a_4}\delta_{a_2a_3}) 
\\
&{}\times \left[
\Db_{F,(2;0)}(123,4) + \Db_{F,(2;0)}(124, 3) 
+ \Db_{F,(2;0)}(134,2) + \Db_{F,(2;0)}(1,234)
\right.
\\
&{}
\left.
\hspace*{.5cm}
- \Db_{F,(2;0)}(12, 34) - \Db_{F,(2;0)}(13,24) - \Db_{F,(2;0)}(14,23)
\right] 
\,.
\end{split}
\ee
The factor $(2 N_c)^{-1}$ appears because we have expressed 
the r.h.s.\ in terms of $\Db_{F,(2;0)}$ instead of $D_{(2;0)}$. 
Interestingly, due to the symmetry of $\Db_{F,(2;0)}$, this additional 
term is completely symmetric in its color and momentum arguments. 
We furthermore observe that due to \eqref{Db20nullstell} it 
vanishes if one of the gluon momenta vanishes, that is for all $i$ we have 
\be
\label{d40addnullstell}
\left.
\Db^{a_1a_2a_3a_4}_{F,(4;0)\,\text{dir}} (\kf_1, \kf_2, \kf_3, \kf_4)
\right|_{\kf_i=0} = 0
\,.
\ee
We will discuss the physical interpretation of this additional piece 
in section \ref{sec:amplitudeseglla} below. 

 \subsubsection{Scalar impact factor}
\label{sec:scalarimpact}

In ${\cal N}=4$ SYM scalars provide a contribution to the 
full impact factor also for larger numbers of gluons. 
A scalar contribution to the six-point function is 
shown in figure~\ref{fig:Scalar6}. 
\FIGURE{
\includegraphics[width=5cm]{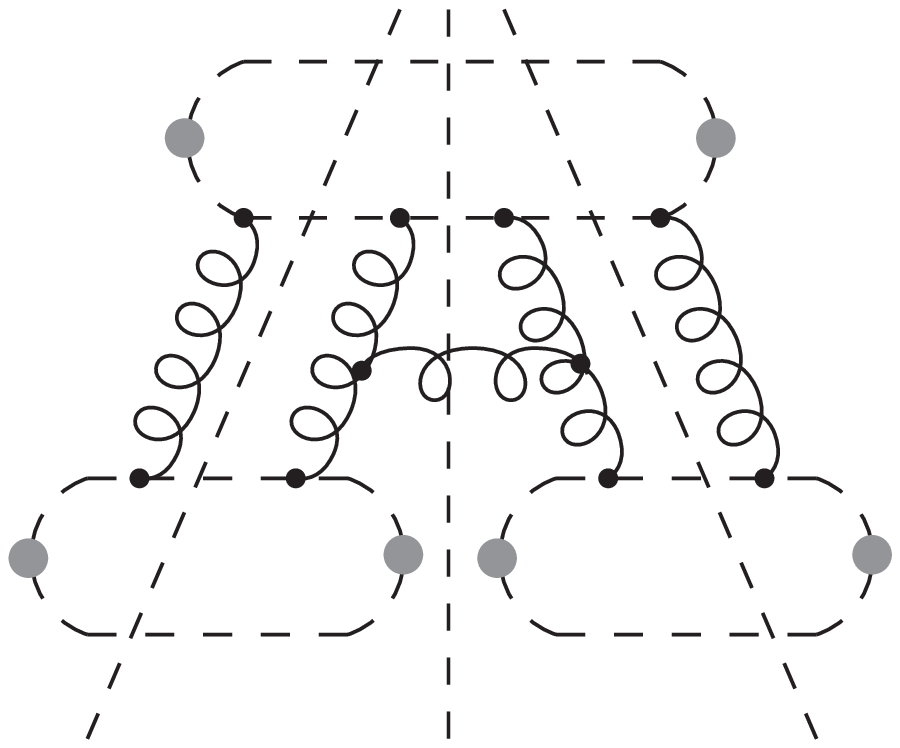}
\caption{A scalar contribution to the six-point function
\label{fig:Scalar6}}
}

In this section we argue that it is possible to relate the scalar impact 
factors for $n$ gluons, $\Db_{S,(n;0)}$, to the two-gluon impact factor 
$\Db_{S,(2;0)}$ in the same way as for fermions. 

Additional diagrams like S5-S9 for the two-point function in 
figure~\ref{fig:IFScalars} also appear with three or more gluons in 
the $t$-channel, see figure~\ref{fig:IFScalars4gluons}. 
But these diagrams are suppressed for the same reasons as in the 
case of two $t$-channel gluons. Every contraction of the gluon 
polarization tensor, 
\begin{equation}
\label{eq:gluontensor}
g^{\mu\nu}=\frac{2}{s}(p^{\mu}_B q^{\nu}_A +
q_A^{\mu} p_B^{\nu})+g^{\mu\nu}_\perp \,,
\end{equation}
with the polarization vector of the $R$-current provides one power of $s$ less 
than in the leading diagrams. Therefore at high energies only diagrams 
with gluons coupled directly to the scalar lines contribute to the 
scalar impact factor.
\FIGURE{
\hspace*{2cm}
  \includegraphics[width=8cm]{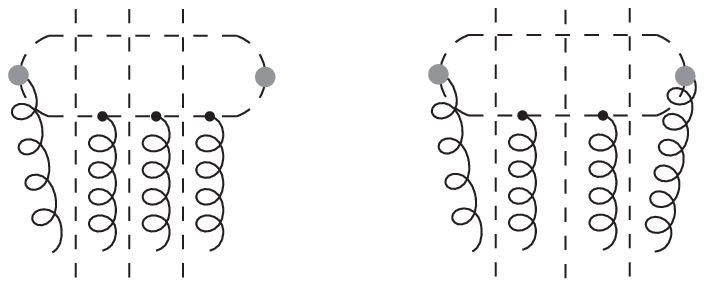}
\hspace*{2cm}
\caption{Two of the additional diagrams in the scalar case
\label{fig:IFScalars4gluons}}
}

Furthermore, diagrams as the one shown in figure~\ref{fig:IFS2gleich} 
do not contribute to the discontinuity that we are considering.%
\FIGURE{
\hspace*{3cm}
\includegraphics[width=3.1cm]{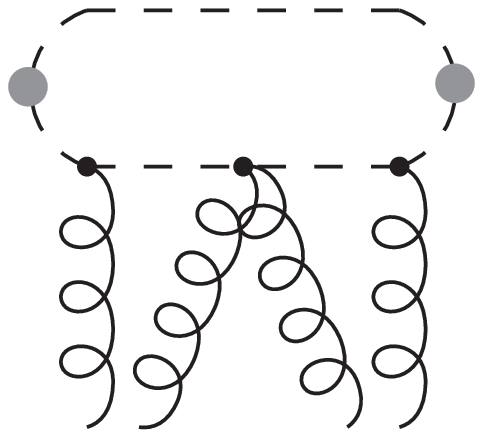}
\hspace*{3cm}
\caption{A further scalar diagram
\label{fig:IFS2gleich}}
}%
Because of the suppression of the additional diagrams also in the
scalar part all impact factors $\Db_{S,(n;0)}$ with arbitrarily many
gluons can be expressed in terms of the two-gluon impact factor
$\Db_{S,(2;0)}$. The reduction mechanism is similar to the one with
fermion loops. The key mechanism for fermions has been explained 
for example in \cite{Bartels:1999aw}, here we will now consider it 
for scalars. 

\FIGURE{
\includegraphics[width=3.5cm]{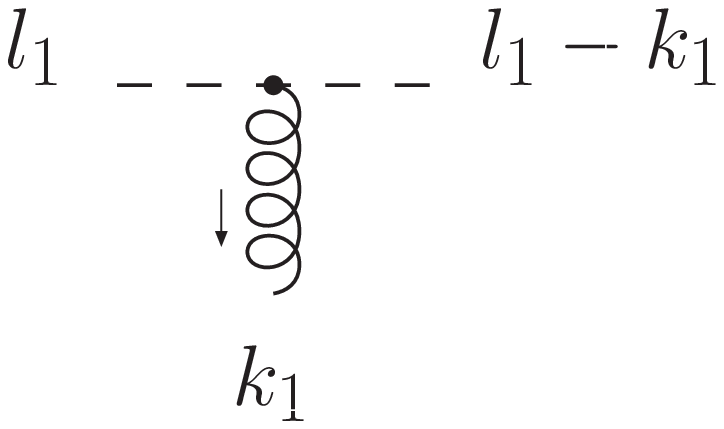}
\caption{Scalar-gluon vertex
\label{fig:Scalarvertex}}
} 
A scalar-gluon vertex, figure~\ref{fig:Scalarvertex}, contracted with the
leading longitudinal part of the gluon
polarization tensor (\ref{eq:gluontensor}), $2 p_B^\mu q_A^\nu/s$, is
proportional to $s \alpha$, where again we make use of a Sudakov
decomposition $l=\alpha q_A+\beta p_B+l_t$ (as in (\ref{eq:Sudakov})) 
of the loop momentum of the scalar loop. 
Let us now consider two adjacent gluons out of the $n$ gluons which
couple to the scalar loop. Every scalar-gluon vertex is contracted 
with the longitudinal momentum $p_B$ from the polarization tensor of 
the $t$-channel gluon. Furthermore, the scalar propagator is on-shell, 
thus resulting in a delta function. Then we obtain
\begin{equation}
\label{eq:sred}
\eqepsres{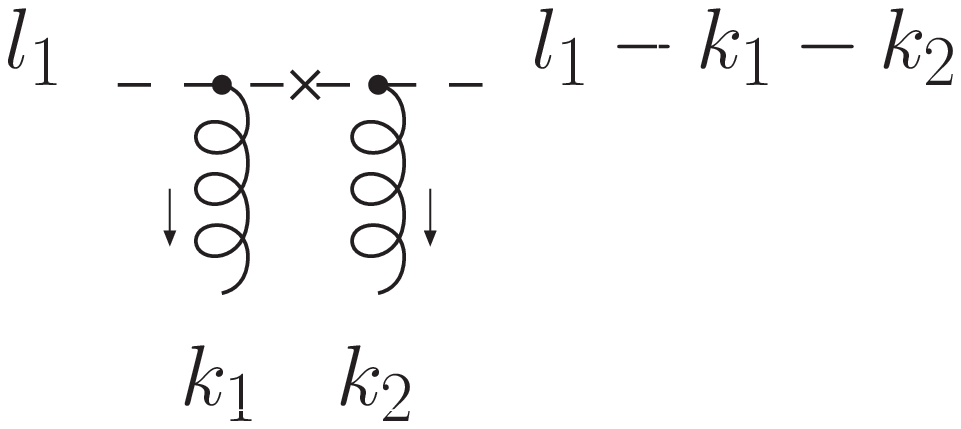}{-4}{4.5cm}  
\hspace*{-.5cm}
\simeq \,\,\, \alpha s (l-k_1 + l - k_1 -k_2) \!\cdot\! p_B \, \,2\pi \delta((l-k_1)^2) \,. 
\end{equation}
Making use of the Sudakov decomposition we find
\begin{equation}
\alpha \,s \,\delta\left((\beta-\beta_{k_1})-{({\bf l}-{\bf k}_1)^2}/{s \alpha }\right)
\,\,\sim
\eqepsres{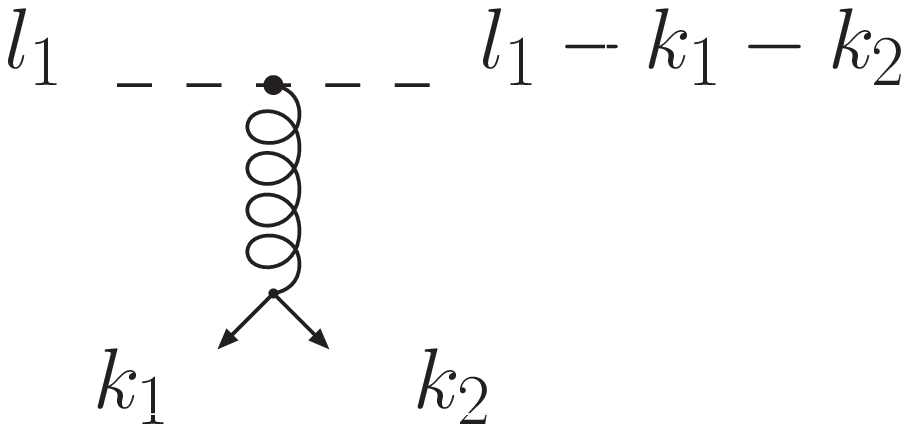}{0}{4.5cm}
\end{equation}
which is the same as the one-gluon-scalar vertex, but with the 
transverse momentum given by the sum of the momenta of both $t$-channel gluons. 
This means that we observe reggeization of the gluons in the scalar part 
of the impact factor as well.

The scalar diagrams contributing to the impact factor come in pairs 
exactly as the fermionic ones. The two diagrams in each pair 
have the same momentum structure 
but the color trace occurs in reversed order. The relative sign between these 
two diagrams is again $(-1)^n$ where $n$ is the number of gluons. 

Altogether, the decomposition of the scalar impact factors $\Db_{S,(n;0)}$ 
into a sum of $\Db_{S,(2;0)}$ works in the same way as for fermions. 
The results are
\bea
  \label{DS30_adj}
  \Db_{S,(3;0)}^{a_1a_2a_3}(\kf_1,\kf_2,\kf_3) 
&=&{} \fr{1}{2} g f_{a_1a_2a_3} \,
[ \Db_{S,(2;0)}(12,3) - \Db_{S,(2;0)}(13,2) + \Db_{S,(2;0)}(1,23) ]  
\eea
and 
\bea
\label{DS40_adj}
\lefteqn{ \Db_{S,(4;0)}^{a_1a_2a_3a_4}(\kf_1,\kf_2,\kf_3,\kf_4)
 \notag } \\
&=& {}   - g^2 d^{a_1a_2a_3a_4} \, [ \Db_{S,(2;0)}(123,4) +
\Db_{S,(2;0)}(1,234) - \Db_{S,(2;0)}(14,23) ]
\nn \\
& & {}   - g^2 d^{a_2a_1a_3a_4} \, [ \Db_{S,(2;0)}(134,2) +
\Db_{S,(2;0)}(124,3) - \Db_{S,(2;0)}(12,34) - \Db_{S,(2;0)}(13,24) ] 
\nn \\
& & {}   + \Db^{ a_1a_2a_3a_4}_{S,(4;0) \,\text{dir}} (1,2,3,4) \,,
\eea 
with $\Db^{ a_1a_2a_3a_4}_{S,(4;0) \,\text{dir}} (1,2,3,4)$ as in 
eq.\ (\ref{D40_add}) but with the index $F$ replaced by $S$ in all terms. 

The full impact factors $\Db_{(3;0)}$ and $\Db_{(4;0)}$ are then given by 
the sum of the fermionic and scalar impact factors as in the two-gluon 
case (\ref{eq:fullimpacttwogluon}):
\be
\begin{split}
\Db_{(3;0)}&= \Db_{F,(3;0)}+\Db_{S,(3;0)} \\
\Db_{(4;0)}&= \Db_{F,(4;0)}+\Db_{S,(4;0)} \,.
\end{split}
\ee
This holds for arbitrary polarization of the $R$-currents.

The key result of the present section is that we have been able to express 
the fermionic parts of the adjoint (${\cal N}=4$) impact factors with up 
to four gluons attached in terms of the corresponding fundamental (QCD) 
impact factors and in terms of one new element, namely the additional 
piece $\Db_{F,(4;0)\, \text{dir}}$. 
The parts that could be related to the fundamental impact factors 
can also be expressed in terms of the adjoint two-gluon impact 
factor $\Db_{F,(2;0)}$ in exactly the same way as the 
fundamental impact factors could be expressed in terms of $D_{(2;0)}$. 
For the scalar contributions to the ${\cal N}=4$ impact factors, 
that is absent in QCD, 
we find an analogous situation. The $n$-gluon impact factors 
$\Db_{S,(n;0)}$ can be expressed in terms of the two-gluon 
impact factor $\Db_{S,(2;0)}$ and in terms of a new element 
$\Db_{S,(4;0)\, \text{dir}}$ which occurs for four gluons. 
The relations in the fermionic and in the scalar sector are 
completely analogous so that they also hold for the 
full impact factors $\Db_{(n;0)}$. 
In section \ref{sec:amplitudeseglla} we will use this observation 
to extract the structure of the solutions to the integral equations in a simple way. 

It is interesting to note that the pattern of reggeization, found for 
$\Db_{(2;0)}$, $\Db_{(3;0)}$, and $\Db_{(4;0)}$, continues for more than 
4 gluons. Similar to $\Db_{(3;0)}$ which, via reggeization, can be 
expressed in terms of a sum of $\Db_{(2;0)}$ impact factors, the five-gluon 
impact factor, $\Db_{(5;0)}$, can be expressed in terms of lower 
impact factors. In particular, we find the new term 
$\Db_{(5;0)\,\text{dir}}$ which can be expressed in terms of the analogous 
four-gluon piece, $\Db_{(4;0)\,\text{dir}}$. More precisely, 
defining 
\be
\label{summeder4dirs}
\Db_{(4;0)\,\text{dir}} = \Db_{F,(4;0)\,\text{dir}} + \Db_{S,(4;0)\,\text{dir}} 
\,,
\ee
we find that 
\begin{eqnarray}
  \label{eq:D50add_byD40add}
\lefteqn{
\Db^{ a_1 a_2 a_3 a_4 a_5}_{(5;0)\,\text{dir}}
(\kf_1, \kf_2, \kf_3, \kf_4, \kf_5)} 
\nn \\
&=&{}\frac{g}{2} 
\left[
f_{a_1a_2c} \,\Db_{(4;0) \text{dir}}^{ca_3a_4a_5}(12,3,4,5) 
+ f_{a_1a_3c} \,\Db_{(4;0)\; \text{dir}}^{ca_2a_4a_5}(13,2,4,5) \right. 
\nn \\
&&{} \hspace{1.2cm} 
+ \,f_{a_1a_4c} \,\Db_{(4;0)\text{dir}}^{ca_2a_3a_5}(14,2,3,5) 
+ f_{a_1a_5c} \,\Db_{(4;0)\,\text{dir}}^{ca_2a_3a_4}(15,2,3,4) 
\nn \\
&&{} \hspace{1.2cm} 
+ \,f_{a_2a_3c} \,\Db_{(4;0)\,\text{dir}}^{a_1ca_4a_5}(1,23,4,5) 
+ f_{a_2a_4c} \,\Db_{(4;0)\,\text{dir}}^{a_1ca_3a_5}(1,24,3,5) 
\nn \\
&&{} \hspace{1.2cm} 
+\, f_{a_2a_5c} \,\Db_{(4;0)\,\text{dir}}^{a_1ca_3a_4}(1,25,3,4) 
+ f_{a_3a_4c} \,\Db_{(4;0)\,\text{dir}}^{a_1a_2ca_5}(1,2,34,5) 
\nn \\
&&{} \hspace{1.2cm} \left.
+ \,f_{a_3a_5c} \,\Db_{(4;0)\,\text{dir}}^{a_1a_2ca_4}(1,2,35,4) 
+ f_{a_4a_5c} \,\Db_{(4;0)\,\text{dir}}^{a_1a_2a_3c}(1,2,3,45)
\right]
\,.
\end{eqnarray}
Further details of the calculation of the five- and six-gluon impact 
factors $\Db_{(5;0)}$ and $\Db_{(6;0)}$ are presented in the appendices 
 \ref{appfive} and \ref{appsix}.

\subsection{Decoupling of the Odderon}
\label{oddsection}

We close this section with an observation that is not along the 
main line of our paper but nevertheless interesting. A closer 
inspection of the color traces that appeared in the 
impact factors considered above allows us to draw some conclusions 
concerning the coupling of the 
Odderon\footnote{For a review on the Odderon, the $C$-odd partner 
of the Pomeron, we refer the reader to \cite{Ewerz:2003xi}.} 
to photon-like impact factors in theories where all particles are the 
adjoint representation of $\text{SU}(N_c)$.

In order to obtain Odderon exchanges in the $t$-channel we 
have to consider impact factors that describe the transition 
from a $C$-odd to a $C$-even state, for example from an 
$R$-current as considered above to a pseudoscalar current. 
As explained in detail in \cite{Braunewell:2005ct} for the 
case of Dirac fermions, the fermionic loop in such an impact factor 
with $n$ gluons attached gives rise to the color factor 
\be
\label{typadjoddcol}
\tr (T^{a_1} \dots T^{a_n}) - (-1)^n \tr (T^{a_n} \dots T^{a_1}) \,.
\ee
In the Pomeron ($C$-even) channel, on the other hand, the impact 
factor containing two vector-like $R$-currents leads to 
\be
\label{typadjoddcol+}
\tr (T^{a_1} \dots T^{a_n}) + (-1)^n \tr (T^{a_n} \dots T^{a_1}) = 2
\tr (T^{a_1} \dots T^{a_n}) \,,
 \ee
as we have seen above. 

In QCD, where the corresponding generators are in the fundamental 
representation, the analog of the above combination of traces 
(\ref{typadjoddcol}) is in general non-zero. 
In particular, the $n$ $t$-gluons form in the Regge-limit a bound state 
for which the EGLLA was formulated in \cite{Braunewell:2005ct}. 
However, if the generators in the trace (\ref{typadjoddcol}) 
are in the adjoint representation, we find
\bea
\label{eq:proofoddtrace}
\tr (T^{a_1} T^{a_2} \dots T^{a_n}) &=&{} (i f_{ka_1l}) (i f_{la_2m})
\dots (i f_{za_nk})
\nn \\
&=&{} (i f_{za_nk}) \dots (i f_{la_2m}) (i f_{ka_1l})
\nn \\
&=&{} (-1)^n (i f_{ka_nz}) \dots (i f_{ma_2l}) (i f_{la_1k})
\nn \\
&=&{} (-1)^n \tr (T^{a_n} \dots T^{a_2} T^{a_1}) \,, 
\eea 
which implies that the combination (\ref{typadjoddcol}) vanishes. As a
consequence, bound states like the Odderon with odd charge parity
decouple from photon-like impact factors, if the particles in the loop
are in the adjoint representation. It should be possible to generalize 
this argument to all possible impact factors in theories that contain 
only particles in the adjoint representation. Odderon contributions 
can occur in such theories only via the splitting of a Pomeron into 
two Odderons \cite{Bartels:1999aw}. Considerations involving the 
direct coupling of the Odderon to particles in a scattering process 
in ${\cal N}=4$ SYM, like for example \cite{Brower:2009zz}, 
therefore require to add particles in the fundamental representation 
at least as external sources. 

\boldmath
\section{Integral equations}
\unboldmath
\label{sec:inteq}

Let us now turn to the integral equations which, in the leading logarithmic 
approximation, sum all graphs contributing to the triple energy 
discontinuity of the six-point function. 
As we will see, the integral 
equations are formally the same in ${\cal N}=4$ SYM and in 
QCD. But the couplings of the 
$t$-channel gluons to the external particles which enter the 
integral equations as initial conditions differ in the two theories, 
as we have discussed in the previous section. 
In order to make a clear distinction between the multi-gluon 
amplitudes in the two theories we will follow the 
notation introduced for the impact factors, i.\,e.\ we will 
denote $n$-gluon amplitudes in QCD 
by normal letters, for example $D_n$, 
while those in ${\cal N}=4$ SYM will be denoted by 
blackboard-style letters, for example $\Db_n$. 

\subsection{Two gluons: BFKL equation}
\label{sec:bfkl}

In the LLA the scattering amplitude is described by the 
well-known BFKL equation \cite{Kuraev:fs,Balitsky:ic} 
which resums all terms of the order $(\alpha_s \log s)^m$. 
For a review of the BFKL equation see for example \cite{Lipatov:1996ts}. 
It is convenient to formulate the BFKL equation and the integral 
equations to be discussed below in Mellin space, that is one trades 
the squared energy $s$ for the complex angular momentum 
$\omega$ by performing a Sommerfeld-Watson transformation. 
(In the following we will suppress the implicit dependence of our 
$n$-gluon amplitudes on $\omega$ in the notation.) 
We consider the amplitude $\Db_2(\kf_1,\kf_2)$ which describes 
the evolution of two reggeized gluons in the $t$-channel, starting 
from an impact factor $\Db_{(2;0)}(\kf_1,\kf_2)$ which couples the two 
gluons to the external currents via a loop of fermions and scalars in the 
adjoint representation. 
The elastic amplitude for $R$-current 
scattering is then obtained from $\Db_2$ by folding it with 
another impact factor $\Db_{(2;0)}$ for the other incoming and 
outgoing $R$-current \cite{Bartels:2008zy}. 

In the LLA the energy dependence of the elastic scattering amplitude 
(from which a corresponding total cross section can be obtained 
via the optical theorem) is then encoded in the BFKL equation 
which we can write for $\Db_2$ as 
\be
\label{bfkl_phi}
\left( \omega - \sum_{i=1}^2\beta(\kf_i)\right) \Db_2^{a_1a_2} = 
\Db_{(2;0)}^{a_1a_2} 
+ K^{ \{b\} \rarr \{a\} }_{2\rarr 2} \otimes \Db_2^{b_1b_2} 
\,. 
\ee
It describes the production of two interacting gluons in the $t$-channel 
with transverse momenta $\kf_1$ and $\kf_2$. The superscripts of 
$\Db_2$ indicate the color labels of the two gluons. 
Graphically, this equation can be illustrated as 
\be
\label{bfkleqdiag}
\left( \omega - \sum_{i=1}^2 \beta(\kf_i)\right) 
\eqeps{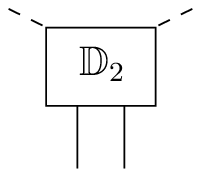}{-2} 
={} 
\eqeps{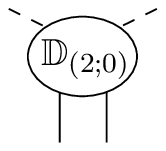}{-1} 
+ \eqeps{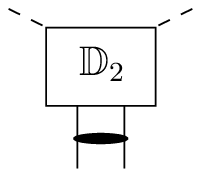}{-2} 
\,.
\ee
The real corrections in the interaction of the gluons are contained in 
the kernel 
\be
\label{2to2kernel}
K^{ \{b\} \rarr \{a\} }_{2\rarr 2} 
(\lf_1,\lf_2;\kf_1,\kf_2) = g^2
 f_{b_1a_1k} f_{ka_2b_2} 
\left[ (\kf_1 +\kf_2)^2 - \frac{\lf_2^2 \kf_1^2}{(\kf_2-\lf_2)^2} 
 - \frac{\lf_1^2 \kf_2^2}{(\kf_1-\lf_1)^2} \right] 
\,, 
\ee
where $g$ is the gauge coupling related to the strong coupling 
constant by $\alpha_s= g^2/(4\pi)$. 
The virtual corrections are given by the gluon trajectory 
function 
\be
\label{traject}
\beta(\kf^2) = 
- \frac{N_c}{2} g^2  \int \frac{d^2\lf}{(2 \pi)^3} 
\frac{\kf^2}{\lf^2 (\lf -\kf)^2} 
\,,
\ee
from which the actual gluon trajectory is obtained as 
$\alpha_g(\kf^2) = 1 + \beta(\kf^2)$. 
The convolution symbol $\otimes$ in eq.\ (\ref{bfkl_phi}) stands 
for a two-dimensional integration over the transverse 
loop momentum $d^2 \lf$ with the measure 
$[(2 \pi)^3 \lf_1^2 \lf_2^2]^{-1}$, and we have the 
condition $\lf_1 + \lf_2 = \kf_1 + \kf_2 = {\bf q}$ with $t = - {\bf q}^2$. 
The inhomogeneous term $\Db_{(2;0)}$ in the BFKL equation (\ref{bfkl_phi}) 
is the impact factor which describes the coupling of the two gluons 
to the scattering particles, for example the coupling of two gluons 
to external $R$-currents through a fermion and scalar loop. 

The sum of the real and virtual contributions to the interaction of the 
two gluons is usually called the BFKL or Lipatov kernel. 
Transforming this kernel to transverse position space one 
finds that it is invariant under M\"obius transformations 
in that two-dimensional space \cite{Lipatov:1985uk}. 
As a consequence it is possible to find the eigenfunctions 
$E^h$ of the kernel in transverse position space and 
to classify them according to their conformal weight $h$. 
The lowest eigenvalue, obtained for $h=1/2$, gives 
rise to the leading behavior $s^{\alpha_{\text{BFKL}}-1}$ of 
the cross section at high energies. 
It is determined by the intercept of the BFKL Pomeron, 
$\alpha_{\text{BFKL}}= 1 + (N_c \alpha_s/\pi) 4 \log 2$. 

In the LLA the BFKL kernel consists of diagrams that 
involve only gluons but no quarks. Only in the NLLA, 
that is if one includes terms of the order $\alpha_s (\alpha_s \log s)^n$, 
quark loops occur in the QCD case \cite{Fadin:1998py,Ciafaloni:1998gs}. 
It is therefore clear that the BFKL kernel in the LLA is not 
affected if we consider a gauge theory with adjoint instead of 
fundamental quarks. In the LLA it is only the impact 
factor $\Db_{(2;0)}$ which differs from the one in QCD with 
fundamental quarks, as we have discussed in section \ref{sec:external}. 
For the NLL corrections to the BFKL equation in ${\cal N}=4$ SYM 
see \cite{Kotikov:2000pm,Kotikov:2002ab}. 

\subsection{More gluons: BKP equations}
\label{sec:glla}

In a straightforward generalization of the BFKL equation, 
referred to as the GLLA, one considers exchanges of more gluons in the 
$t$-channel, still keeping the number $n$ of gluons fixed during 
the evolution. The corresponding $n$-gluon amplitudes 
are described by the BKP 
equations \cite{Bartels:1980pe,Kwiecinski:1980wb} 
which resum terms containing the maximally possible 
number of logarithms for a given fixed $n$. 
These BKP states appear in several places (see further below), and 
we briefly summarize their most important properties. 
In leading order, this equation can be written completely in terms of the 
gluon trajectory function $\beta$ and of the kernel 
$K_{2 \to 2}$ of \eqref{2to2kernel} 
that occurred already in the BFKL equation. 
It is therefore immediately clear that, at the leading logarithmic level, 
also the BKP equation is identical in QCD with fundamental quarks and in 
${\cal N}=4$ SYM. 

Let us consider amplitudes $\Bb_n$ describing the production of 
$n$ gluons in the $t$-channel, similar to the BFKL amplitude $\Db_2$ 
above. 
The BKP equation for the amplitude $\Bb_n$ in ${\cal N}=4$ SYM 
reads 
\bea
\label{BKPn_adjoint}
\left( \omega - \sum_{i=1}^n \beta(\kf_i) \right) \Bb_n^{a_1\dots a_n} 
&=& {}
\Bb_{(n;0)}^{a_1 \dots a_n} 
+ \sum K^{ \{b\} \rarr \{a\} }_{2\rarr 2} \otimes \Bb_n^{b_1 \dots b_4}
\,,
\eea
where now the amplitude carries $n$ color labels $a_i$. 
Again the evolution starts with some initial condition, $\Bb_{(n;0)}$, 
given by an impact factor or by a transition from another multigluon state 
(see below). 
The sum extends over all pairwise interactions of the $n$ gluons. 
The kernel (\ref{2to2kernel}) has to be interpreted 
such that only the two gluons participating in the respective 
interaction enter the kernel, while the other $n-2$ gluons 
do not change their color nor their momentum. 
We can illustrate the BKP equation graphically (for the example $n=4$) 
as 
\be
\label{inteqB4diag}
\left( \omega - \sum_{i=1}^4 \beta(\kf_i)\right) 
\eqeps{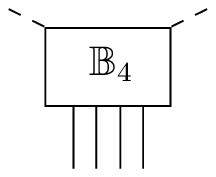}{-2} 
= 
\eqeps{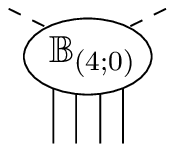}{-1} 
+ \sum \eqeps{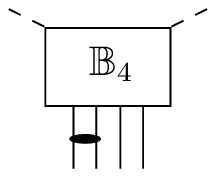}{-2} 
\,.
\ee
Here the BKP equation is written for amplitudes $\Bb_n$ with 
multiple cuts, that is one takes $n-1$ discontinuities 
corresponding to the energy variables defined from the 
four-momentum of the incoming photon and those of 
the first $i$ gluons, $s_i=(q + \sum_{j=1}^1 k_j)^2$, 
with $1\le i \le n-1$. As a consequence of this the $t$-channel 
gluons do not cross each other and the $s$-channel gluons 
exchanged between them via the kernels $K_{2 \to 2}$ are 
on-shell. 

In NLLA the BKP equation will 
be different for QCD and ${\cal N}=4$ SYM. The fermionic 
contribution to the NLL terms can be 
obtained by replacing the gluon trajectory and the real part $K_{2 \to 2}$
of the BFKL kernel by the corresponding next-to-leading QCD 
expressions, with the quarks taken in the representation under 
consideration. However, in addition also a new kernel will appear, 
$K_{3 \to 3}$, which has not been computed yet. 

Let us recall here that in the large-$N_c$ limit the Hamiltonian 
corresponding to the BKP equation is integrable \cite{Lipatov:1993yb}. 
More precisely, it is equivalent to the 
XXX-Heisenberg model of $\mbox{SL}(2,\C)$ spin zero 
\cite{Lipatov:1994xy,Faddeev:1994zg}. At large $N_c$ 
the leading terms in the BKP equation are those in which 
the integral kernel acts only on neighboring gluons $i$ and $i+1$ 
with periodic boundary conditions. The color structure 
simplifies such that each pair of neighboring gluons is 
in a color octet state. 
For the six-point $R$-current correlator, however, we will show that 
the large-$N_c$ limit suppresses the BKP states. For the rest of the paper 
we will stay with finite $N_c$, and only at the end we will 
consider the large-$N_c$ limit. 

\subsection{Changing the number of gluons: coupled 
integral equations}
\label{sec:eglla}

In order to compute the sum of diagrams that contribute to the 
six-point function (examples have been given in figure~\ref{fig:diagrams}) 
we need to couple amplitudes with 2, 3, and 4 gluons,
$\Db_{(2;0)}^{a_1a_2}$, $\Db_{(3;0)}^{a_1a_2a_3}$, 
and $\Db_{(4;0)}^{a_1a_2a_3a_4}$.
It is beyond the scope of the present paper to describe in full detail 
how the relevant $n$-gluon amplitudes $\Db_n$ are obtained 
from multi-particle scattering amplitudes by applying suitable 
discontinuities similar to the ones mentioned in the previous 
section below eq.\ \eqref{inteqB4diag}. 
A more detailed account of this procedure has 
been given in \cite{Bartels:1994jj} for the example of the four-gluon 
amplitude $D_4$. 

In the following we simply write down the integral equations 
for the $n$-gluon amplitudes which arise in the so-called 
extended GLLA, or EGLLA. This approximation scheme can be 
characterized as resumming the maximally possible number 
of logarithms for a given number of $t$-channel gluons $n$ 
the production of which is described by an amplitude $\Db_n$, 
with number-changing transitions being allowed. For the details 
of the EGLLA we refer the reader to \cite{Bartels:1999aw,Bartels:unp}. 
Various aspects of the EGLLA have been studied in 
\cite{Lotter:1996vk}-\cite{Bartels:2008ru}. 

Let us now consider the integral equations of the EGLLA for the 
amplitudes $\Db_n$ for ${\cal N}=4$ SYM 
and discuss the elements entering them in more detail. 
For up to $n=4$ gluons in the $t$-channel these coupled integral 
equations read 
\be
\label{inteq2_adjoint}
\left( \omega - \sum_{i=1}^2\beta(\kf_i)\right) \Db_2^{a_1a_2} = 
\Db_{(2;0)}^{a_1a_2} 
+ K^{ \{b\} \rarr \{a\} }_{2\rarr 2} \otimes \Db_2^{b_1b_2} 
\ee
\be
\label{inteq3_adjoint}
\left( \omega - \sum_{i=1}^3 \beta(\kf_i) \right) \Db_3^{a_1a_2a_3} = 
\Db_{(3;0)}^{a_1a_2a_3} 
+ K^{ \{b\} \rarr \{a\} }_{2\rarr 3} \otimes \Db_2^{b_1b_2} 
+ \sum K^{ \{b\} \rarr \{a\} }_{2\rarr 2} \otimes \Db_3^{b_1b_2b_3}
\ee
\bea
\label{inteq4_adjoint}
\left( \omega - \sum_{i=1}^4 \beta(\kf_i) \right) \Db_4^{a_1a_2a_3a_4} 
&=& {}
\Db_{(4;0)}^{a_1a_2a_3a_4} 
+ K^{ \{b\} \rarr \{a\} }_{2\rarr 4} \otimes \Db_2^{b_1b_2} 
+ \sum K^{ \{b\} \rarr \{a\} }_{2\rarr 3} \otimes \Db_3^{b_1b_2b_3} \nn \\
&&{} + \sum K^{ \{b\} \rarr \{a\} }_{2\rarr 2} \otimes \Db_4^{b_1b_2b_3b_4}
\,.
\eea
As in the case of the BKP equations the kernels 
$K^{ \{b\} \rarr \{a\} }_{2\rarr m}$ have to be interpreted in such 
a way that only two gluons undergo an interaction resulting in $m$ 
gluons, while the other gluons of the amplitude to which the kernel is 
applied keep their momentum and color. 
It is straightforward to obtain the integral equations for the 
higher $n$-gluon amplitudes following the above pattern. 
In appendix \ref{appfive} we discuss the case $n=5$. 

The lowest order terms $\Db_{(n;0)}$ describe the coupling of $n$ 
gluons to the $R$-currents, as discussed in 
detail in section \ref{sec:external} above. 
The trajectory function $\beta$ is the one given in eq.\ (\ref{traject}), 
while we define the transition kernel from 2 to $m$ gluons that occurs 
in the integral equations as 
\bea
\label{2tomkernel}
\lefteqn{K^{ \{b\} \rarr \{a\} }_{2\rarr m} 
(\lf_1,\lf_2;\kf_1, \dots, \kf_m) }
\nn \\ 
&=&{}
g^m f_{b_1a_1k_1} f_{k_1a_2k_2}\dots f_{k_{m-1}a_mb_2} 
\left[ (\kf_1 + \dots + \kf_m)^2 
- \frac{\lf_2^2 (\kf_1 + \dots + \kf_{m-1})^2}{(\kf_m-\lf_2)^2} 
\right.
\nn \\
&&{}
\left.
- \frac{\lf_1^2 (\kf_2 + \dots + \kf_m)^2}{(\kf_1-\lf_1)^2} 
+ \frac{\lf_1^2 \lf_2^2 
(\kf_2 + \dots + \kf_{m-1})^2}{(\kf_1-\lf_1)^2 (\kf_m - \lf_2)^2}
\right] \,,
\eea
where the $f_{klm}$ are again the structure constants of $\text{su}(N_c)$. 
Note that for $m=2$ the definition (\ref{2tomkernel}) 
reduces\footnote{In this case the last term in square brackets 
in eq.\ (\ref{2tomkernel}) is understood to vanish by definition.} 
to the kernel $K_{2 \to 2}$ of (\ref{2to2kernel}) 
which occurred in the BFKL and in the BKP equations. 
We stress that these kernels do not depend on the 
color representation of the quarks, and are therefore exactly the same as 
in QCD with fundamental quarks. As in the BFKL equation, fermionic 
and scalar contributions enter the kernels only at the next-to-leading level. 

As in the BFKL and BKP equations the convolution symbol 
$\otimes$ stands for an integration over the loop momentum 
with the measure $[(2 \pi)^3 \lf_1^2 \lf_2^2]^{-1}$, and here 
we have the condition that $\lf_1 + \lf_2 = \kf_1 + \dots + \kf_m$ 
in \eqref{2tomkernel}. 
We write the kernel (\ref{2tomkernel}) diagrammatically as 
\be
\label{diagkernel2m}
\picresize{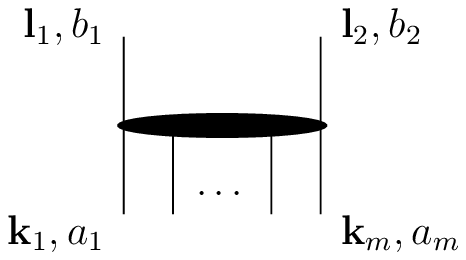}{4cm} 
\,. 
\ee

With the help of the diagrammatic representation (\ref{diagkernel2m}) 
of the kernels we can write the hierarchy of coupled integral equations of 
the EGLLA in a more intuitive diagrammatic form as 
\bea
\label{inteqD2diag}
\left( \omega - \sum_{i=1}^2 \beta(\kf_i)\right) 
\eqeps{glD21}{-2} 
&=&{} 
\eqeps{glD22}{-1} 
+ \eqeps{glD23}{-2} 
\\
\label{inteqD3diag}
\left( \omega - \sum_{i=1}^3 \beta(\kf_i)\right) 
\eqeps{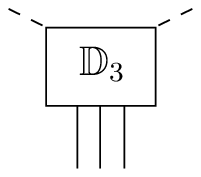}{-2} 
&=&{} 
\eqeps{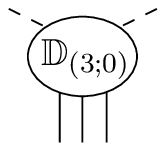}{-1} 
+ \eqeps{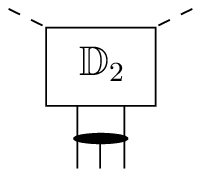}{-2} 
+ \sum \eqeps{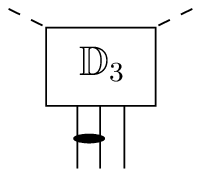}{-2} 
\\
\label{inteqD4diag}
\left( \omega - \sum_{i=1}^4 \beta(\kf_i)\right) 
\eqeps{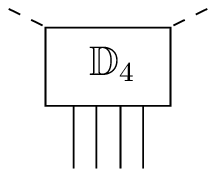}{-2} 
&=&{} 
\eqeps{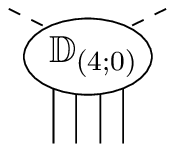}{-1} 
+ \eqeps{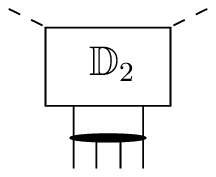}{-2} 
+ \sum \eqeps{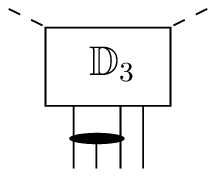}{-2} 
\nn \\
&&{}
+ \sum \eqeps{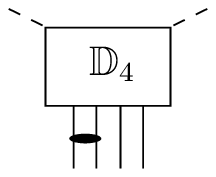}{-2} 
\,
\eea
The sums in the integral equations extend over all possible permutations 
of the gluon lines in the $t$-channel under the condition that these 
lines do not cross each other. For a more detailed description of this 
point including explicit examples we refer the reader to 
\cite{Bartels:1999aw}. Note that in each diagram only two of the 
gluons enter the interaction defined by the kernel. 
The momenta and color labels of the other gluons are not affected 
by the kernel. 

The integral equations eqs.\ (\ref{inteq2_adjoint})-(\ref{inteq4_adjoint}) 
form the basis of our analysis in the following sections. Our strategy will 
be to relate certain parts of the amplitudes for the case of ${\cal N}=4$ SYM 
to the amplitudes of QCD with fundamental quarks, and to invoke 
known results of the QCD case. 
The difference between the amplitudes for the two theories clearly 
originates from the different impact factors. 

\boldmath
\section{Solutions for $\Db_3$ and $\Db_4$}
\unboldmath
\label{sec:amplitudeseglla}

In order to find solutions for the amplitudes $\Db_3$ and $\Db_4$ 
we will make use of the results obtained in section 
\ref{sec:external} for the impact factors $\Db_{(n;0)}$.
In the case of two gluons, the amplitude $\Db_2$ is given by the 
BFKL Pomeron Green's function, convoluted with the two-gluon 
impact factor $\Db_{(2;0)}$. 
In particular, for the fermionic part $\Db_{F,2}$, that is the one with 
only a fermionic loop in the impact factor, we have the simple relation
\be
\label{bfkladj-bfklfund}
\Db_{F,2} (\kf_1, \kf_2) = 2 N_c \,R \,D_2 (\kf_1, \kf_2)
\,.
\ee
which is an immediate consequence of 
the fact that the BFKL equation with a fundamental 
impact factor differs from eq.\ \eqref{bfkl_phi} only by 
a relative factor in the impact factor. 

Considering then the three-gluon amplitude $\Db_3$, we 
observe that the three-gluon impact factor $\Db_{(3;0)}$ is related 
to the two-gluon impact factor $\Db_{(2;0)}$ in exactly 
the same way as $D_{(3;0)}$ was related to $D_{(2;0)}$ in 
QCD with fundamental quarks, see eqs.\ \eqref{D30_fund} and 
\eqref{D30_adj}. One can then easily verify that the solution 
to the integral equation \eqref{inteq3_adjoint} is given by 
\be
\label{d3_adjoint}
\Db_3^{a_1a_2a_3}(\kf_1,\kf_2,\kf_3) 
= \fr{1}{2} g f_{a_1a_2a_3} \,
[ \Db_2(12,3) - \Db_2(13,2) + \Db_2(1,23) ]  
\,,
\ee
in complete analogy to the case of QCD with fundamental quarks 
treated in \cite{Bartels:1992ym,Bartels:1994jj}. (Throughout this 
section we again make use of the notation \eqref{notation}.) 
We emphasize that this result holds for the complete amplitude 
with the impact factors containing now the sum of fermionic 
and scalar loops. 
Note that also in the full amplitude we observe reggeization. The 
amplitude $\Db_3$ is a superposition of two-gluon amplitudes 
$\Db_2$ in each of which one gluon is composed of two gluons at 
the same position 
in transverse space (or, equivalently, the amplitude depends only 
on the sum of their transverse momenta). In other words, such 
a pair of gluons at the same point in transverse space acts like a 
single gluon, in this way forming a so-called reggeized gluon. 
One can actually show that a self-consistent 
solution is found only if all gluons exchanged in the $t$-channel 
are reggeized. What we observe in the amplitude $\Db_3$ is a 
perturbative expansion of the reggeized gluon in terms of 
more `elementary gluons' to the first non-trivial order, or in other 
words a higher (two-particle) Fock state of the reggeized gluon. 
For a more detailed discussion of reggeization in the EGLLA 
see \cite{Ewerz:2001fb}. 

While in the three-gluon case the full amplitude has the same 
color structure and dependence on the gluon momenta as 
the impact factor (compare eqs.\ \eqref{D30_fund} and 
\eqref{d3_adjoint}) the structure becomes more interesting 
already when we consider four gluons. 
We have found in section \ref{sec:external} that in ${\cal N}=4$ SYM 
the impact factor with four gluons $\Db_{(4;0)}$ consists of two parts 
exhibited in \eqref{D40_adj} and \eqref{DS40_adj}: 
one which can be expressed in terms of the adjoint two-gluon impact 
factor $\Db_{(2;0)}$ in exactly the same way as 
the fundamental four-gluon impact factor $D_{(4;0)}$ was 
expressed in terms of $D_{(2;0)}$ (compare eqs.\ \eqref{D40_adj} and 
\eqref{DS40_adj} to \eqref{d40}), and a new additional part 
$\Db_{(4;0) \,\text{dir}}$, see eq.\ \eqref{summeder4dirs}. 
Both parts have a fermionic contribution and a scalar contribution. 
The similarity of the first part to the case of QCD with fundamental quarks 
suggests to attack the integral equation \eqref{inteq4_adjoint} 
for $\Db_4$ in the same 
way as was done in \cite{Bartels:1992ym, Bartels:1993ih, Bartels:1994jj} 
for the case of $D_4$. We hence split $\Db_4$ into two parts 
\be
\label{d4risplit}
\Db_4 = \Db_4^R + \tilde{\Db}_4\,,
\ee 
and make an ansatz for the so-called reggeizing part $\Db_4^R$. 
Similarly to the case of three gluons above this ansatz is obtained 
from the first parts of the impact factors \eqref{D40_adj} and 
\eqref{DS40_adj} by 
replacing the lowest order terms $\Db_{(2;0)}$ by full amplitudes 
$\Db_{F,2}$ and $\Db_{S,2}$ while keeping the color and momentum structure. 
Adding the fermionic and scalar contributions we have 
\bea
\label{d4r_adjoint}
\lefteqn{\Db_4^{R\,a_1a_2a_3a_4}(\kf_1,\kf_2,\kf_3,\kf_4)} \nn \\
  &=&{}  - g^2 d^{a_1a_2a_3a_4} \, [ \Db_2(123,4) + \Db_2(1,234) 
                                  - \Db_2(14,23) ] 
   \nn \\
 & &{} - g^2 d^{a_2a_1a_3a_4}  \, [ \Db_2(134,2) + \Db_2(124,3)
                                 - \Db_2(12,34) - \Db_2(13,24) ] \,.
\eea
We will discuss its reggeizing structure later, see 
the discussion of eq.\ \eqref{Db4soldiag} below. 
Inserting now \eqref{d4risplit} with \eqref{d4r_adjoint} in the integral 
equation \eqref{inteq4_adjoint} one can derive a new integral 
equation for the remaining part $\tilde{\Db}_4$. The procedure for 
this is exactly the same as in the case of fundamental quarks (for 
a detailed description we refer the reader to \cite{Bartels:1999aw}). 
This complete analogy allows us to use the results of 
\cite{Bartels:1992ym, Bartels:1993ih, Bartels:1994jj}. We find 
that $\tilde{\Db}_4$ satisfies the integral equation 
\bea
\label{inteq4tilde_adjoint}
  \left(\omega - \sum_{i=1}^4 \beta(\kf_i) \right)
  \tilde{\Db}_4^{a_1a_2a_3a_4} (\kf_1,\kf_2,\kf_3,\kf_4) 
&=&{} 
   V_{2 \rightarrow 4}^{a_1a_2a_3a_4} \Db_2
   + \Db_{(4;0)\, \text{dir}}^{a_1a_2a_3a_4}
\nn \\
&&{}
+\sum K^{ \{b\} \rarr \{a\} }_{2\rarr 2} \otimes \tilde{\Db}_4^{b_1b_2b_3b_4}
   \,,
\eea
where $V_{2 \rightarrow 4}$ is exactly the same two-to-four gluon 
transition vertex which was derived in \cite{Bartels:1994jj}. 
$V_{2 \rightarrow 4}$ acts as an integral operator on the two-gluon 
amplitude $\Db_2$ and thus couples it to the four-gluon BKP state. 
The vertex can be written in terms of an infrared-finite function $V$ as 
\bea
\label{colV}
 V_{2 \rightarrow 4}^{a_1a_2a_3a_4}(\{ {\bf q}_j\},\kf_1,\kf_2,\kf_3,\kf_4) 
 &=&{} 
  \delta_{a_1a_2} \delta_{a_3a_4} 
V(\{ {\bf q}_j\},\kf_1,\kf_2;\kf_3,\kf_4)
 \nonumber \\
 & &{} 
 +\, \delta_{a_1a_3} \delta_{a_2a_4} 
V(\{ {\bf q}_j\},\kf_1,\kf_3;\kf_2,\kf_4)
 \nonumber \\
 & &{}
 +\,\delta_{a_1a_4} \delta_{a_2a_3} 
V(\{ {\bf q}_j\},\kf_1,\kf_4;\kf_2,\kf_3)
\,,
\eea
where the ${\bf q}_j$ are the transverse momenta of the two 
gluons in the amplitude $\Db_2$. 
The vertex $V_{2\to 4}$ 
is hence completely symmetric in the four outgoing gluons, that 
is under the simultaneous exchange of their color labels and momenta. 
An important property of this transition vertex is that upon Fourier 
transformation to two-dimensional impact parameter space it 
is invariant under conformal (M\"obius) transformations in the 
transverse plane \cite{Bartels:1995kf}. Another important property 
of the two-to-four gluon vertex is that it vanishes if the transverse 
momentum of any of the outgoing gluons vanishes. Further properties of 
$V_{2 \rightarrow 4}$ have been found and discussed in 
\cite{Braun:1997nu,Ewerz:1999gk,Ewerz:2001uq,Schatz:2003rr}. 

Note that the new integral equation \eqref{inteq4tilde_adjoint} 
for $\tilde{\Db}_4$ is a linear integral equation involving only the 
integral kernel $\sum K_{2 \to 2}$ of the four-gluon BKP equation, 
compare eq.\ \eqref{BKPn_adjoint}. The terms involving 
$K_{2\to 4}\otimes \Db_2$ and $K_{2\to 3}\otimes \Db_3$ 
present in the original integral equation \eqref{inteq4_adjoint} 
have disappeared in favor of $V_{2 \to 4}\otimes \Db_2$. The latter 
has now become a contribution to the inhomogeneous term in the linear 
equation for $\tilde{\Db}_4$. Since the equation \eqref{inteq4tilde_adjoint} 
is linear and its inhomogeneous term is a sum of two terms we can 
write its solution as a sum, 
\be
\label{splitDbtilde}
\tilde{\Db}_4 = \Db_4^I + \Db_{4\,\text{dir}}
\,,
\ee
with the two parts satisfying 
\be
\label{inteqDb4I}
  \left(\omega - \sum_{i=1}^4 \beta(\kf_i) \right)
  \Db_4^{I\,a_1a_2a_3a_4} (\kf_1,\kf_2,\kf_3,\kf_4) =
   V_{2 \rightarrow 4}^{a_1a_2a_3a_4} \Db_2
  +\sum K^{ \{b\} \rarr \{a\} }_{2\rarr 2} \otimes \Db_4^{I\,b_1b_2b_3b_4}
\ee
and 
\be
\label{inteqDb4dir}
 \left(\omega - \sum_{i=1}^4 \beta(\kf_i) \right)
  \Db_{4\,\text{dir}}^{a_1a_2a_3a_4} (\kf_1,\kf_2,\kf_3,\kf_4) = 
  \Db_{(4;0)\, \text{dir}}^{a_1a_2a_3a_4}
  +\sum K^{ \{b\} \rarr \{a\} }_{2\rarr 2} \otimes \Db_{4\,\text{dir}}^{b_1b_2b_3b_4}
   \,,
\ee
respectively. In both equations the inhomogeneous terms are known 
so that the solution can formally be obtained by iteration of the integral 
kernel $\sum K_{2\to 2}$. 

The amplitude $\Db_4^I$ in eq.\ \eqref{inteqDb4I} is well known from 
the four-gluon amplitude in QCD with fundamental quarks. 
More precisely, there one can split the amplitude $D_4$ according 
to $D_4=D_4^R + D_4^I$, and $D_4^I$ satisfies the same equation 
\eqref{inteqDb4I} as $\Db_4^I$ here. The two amplitudes are in 
fact proportional to each other, 
$\Db_4^I = 2 N_c \,R D_4^I$, and the relative factor 
originates only from the impact factor. As one can see from the 
iterative solution of eq.\ \eqref{inteqDb4I}, $\Db_4^I$ 
contains a two-gluon state which couples to the impact factor and then at 
some point undergoes a transition to a four-gluon 
state. The vertex $V_{2 \rightarrow 4}$ describing this transition 
is local in rapidity. 

The last part $\Db_{4\,\text{dir}}$ of the four-gluon amplitude
$\Db_4$ is new in the supersymmetric theory and originates from the
fact that particles inside the loop are in the adjoint representation.
It has a simple structure as can be read off from the integral
equation \eqref{inteqDb4dir}: it consists of a four-gluon BKP state
that is directly coupled to the loop of adjoint fermions and scalars,
and the coupling is just given by the additional term
$\Db_{(4;0)\,\text{dir}}$ of eq.\ \eqref{D40_add} and the scalar analog 
of that equation. For the case of five
$t$-channel gluons, addressed in appendix \ref{appfive}, it
can be further shown, that the additional term that arises there due
to the adjoint representation reggeizes in terms of
$\Db_{4\,\text{dir}}$, in the very same way as $\Db_{3 }$ reggeizes in
terms of $\Db_{2 }$.

In summary, we have decomposed the four-gluon amplitude $\Db_4$ 
into three parts, 
\be
\label{split_d4_adjoint}
\Db_4 = \Db_4^R + \Db_4^I + \Db_{4\, \text{dir}} \,, 
\ee
and have been able to derive their structure\footnote{In 
\cite{ Bartels:2009ms, Bartels:2009zc} it has been found that for the 
large-$N_c$ expansion of the six-point amplitude in topologies of 
two-dimensional surfaces the above decomposition occurs automatically, 
as a consequence of different classes of color structures.}. 
We can illustrate these three parts graphically in the following way: 
\be
\label{Db4soldiag}
\Db_4 = \sum \eqepsres{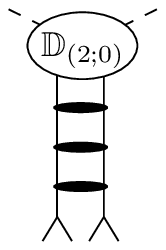}{-2}{1.8cm} 
+ \eqepsres{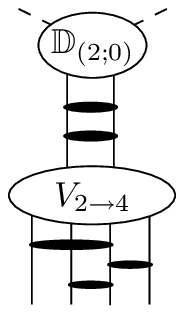}{-2}{2cm} + \eqepsres{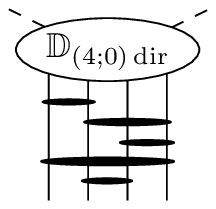}{-2}{2.3cm} 
\,.
\ee
The first term $\Db_4^R$, which we have not yet discussed in detail here, 
is again a reggeizing term, namely 
a superposition of two-gluon amplitudes $\Db_2$. The sum extends 
over all possible (that is seven) partitions of the four gluons into 
two non-empty sets, see eq.\ \eqref{d4r_adjoint}. In each of these terms several 
gluons are at the same point in transverse space and form a reggeized 
gluon that carries the sum of their transverse momenta. 
As in the case of the three-gluon amplitude $\Db_3$ \eqref{d3_adjoint} 
we can interpret these composite gluons as representing higher Fock 
states of the reggeized gluon. An interesting aspect of reggeization is the 
color structure that is associated with this process. Recall that in 
the three-gluon amplitude $\Db_3$ \eqref{d3_adjoint} this combination 
of two gluons into a more composite gluon always came with a 
$f$-structure constant of the $\text{SU}(N_c)$ gauge group. 
In the reggeizing part $\Db_4^R$ of the four-gluon amplitude we 
see also other contributions, namely a $d$-tensor of type \eqref{d_abcd} 
for the combination of three gluons into one, and also $d$-type symmetric 
structure constants and even $\delta$-tensors for the combination of 
two gluons into one. The latter are obtained from the decomposition 
\eqref{decomposedtens} of the $d^{abcd}$-tensor given in appendix 
\ref{sec:color}. 
A more detailed discussion of these color tensors 
in the context of reggeization was given in \cite{Ewerz:2001fb}. 

In the second term in eq.\ \eqref{Db4soldiag}, $\Db_4^I$, 
two gluons couple to the impact factor, 
interact according to BFKL evolution, then undergo a transition 
to four gluons via the vertex $V_{2 \to 4}$, and finally these four gluons 
interact pairwise according to BKP evolution. In the third term, 
$\Db_{4\, \text{dir}}$, four gluons couple directly to the 
fermion and scalar loop 
and then interact pairwise according to BKP evolution, 
without a transition vertex in the evolution. 
Both the first and second term are present already in QCD with 
fundamental quarks. In ${\cal N}=4$ SYM 
only their normalization is different, being $(2 N_c R)$ times 
that of the corresponding terms for QCD with fundamental quarks. 
The third term occurs only in theories with adjoint particles, 
here fermions and scalars in ${\cal N}=4$ SYM. 

A crucial step towards a better understanding of the $n$-gluon 
amplitudes $D_n$ in QCD was 
the observation that they exhibit the structure of a field theory 
of gluon exchanges, and the same observation applies to our 
results for ${\cal N}=4$ SYM. 
There are $n$-gluon states with fixed numbers of reggeized gluons in 
the $t$-channel, for example the two-gluon state and the four-gluon 
state. In addition there are transition vertices coupling those 
states to each other, like for example the two-to-four gluon 
transition vertex $V_{2 \to 4}$. This structure has been 
verified in the explicit calculation of the amplitudes with up 
to six gluons \cite{Bartels:1999aw}. In the six-gluon case a 
new transition vertex from two to six gluons occurs which 
contains a Pomeron-Odderon-Odderon coupling as well as 
a one-to-three Pomeron transition. In the Pomeron channel 
it turns out that only $n$-gluon states with even numbers $n$ of 
gluons occur. The amplitudes $D_n$ with odd $n$ reggeize 
and are hence superpositions of amplitudes with 
less and even numbers of gluons. 
(In the appendix we show that this pattern 
continues for 5 gluons also in ${\cal N}=4$ SYM.) 

The amplitudes in the Pomeron channel hence consist of only 
very few elements, namely states of even numbers of reggeized 
gluons and transition vertices between them. All of these 
elements of the field theory of reggeized gluon exchanges possess 
two important properties: firstly, they are completely symmetric 
in the exchange of any two gluons, and secondly, they vanish 
when any of the transverse gluon momenta is set to zero. 
This property holds for the two-to-four gluon vertex $V_{2 \to 4}$, 
as we have pointed out above, and it also holds for the two-to-six 
gluon transition vertex \cite{Bartels:1999aw}. The same 
holds also for the $n$-gluon states, once they are coupled to 
an element that fulfills these two conditions. It is therefore 
plausible to regard also the impact factor with two-gluons 
$D_{(2;0)}$ as a fundamental element of our field theory. 
As we have pointed out in section \ref{sec:external} it 
in fact satisfies these conditions. Moreover, all higher impact 
factors with fundamental quarks could be expressed as 
superpositions of $D_{(2;0)}$, thus exhibiting reggeization, 
see again section \ref{sec:external}. 

Furthermore, and from a theoretical point of view most 
importantly, all elements of the field theory are conformally 
invariant, that is they are invariant under M\"obius transformations 
of the gluon coordinates in two-dimensional impact parameter space. 
Hence the complete amplitudes exhibit the structure of 
an effective conformal field theory of reggeized gluon exchanges 
at high energy. 

Let us now return to ${\cal N}=4$ SYM and let us 
reconsider our results in the light of their interpretation in the framework of 
such an effective conformal field theory. The two- and the three-gluon 
amplitudes behave as the corresponding amplitudes in QCD 
with fundamental quarks and hence share their properties regarding 
the field theory structure. In particular we have found the absence 
of an actual three-gluon state in $\Db_3$ due to reggeization. 
The first two terms of the four-gluon amplitude $\Db_4$ in 
eq.\ \eqref{Db4soldiag} reproduce exactly the structure of the amplitude 
$D_4$ in QCD with fundamental quarks, and hence have the same 
field theory structure discussed above. The new term 
$\Db_{4\, \text{dir}}$ contains again the four-gluon state in 
the $t$-channel with properties that fit the field theory structure. 
But in addition it contains as a new element the direct coupling 
$\Db_{(4;0)\, \text{dir}}$ of the four gluons to the impact factor consisting of 
adjoint fermions and scalars. Inspection of its explicit structure 
quickly shows that it is symmetric under the exchange of any 
two gluons, and that it vanishes if one of the gluon momenta 
vanishes, see eq.\ \eqref{d40addnullstell}. Therefore the new contribution 
to the four-gluon impact factor satisfies the expectations that we 
have for a new element of our field theory. 

\section{The six-point amplitude}
\label{sec:partialwave}

Collecting the results of the previous sections, we now return to the
six-point amplitude defined in section \ref{sec:sixpoint}. As we have
explained before, the partial wave $F(\omega_1,\omega_2,\omega;
t_1,t_2,t)$ is obtained as a convolution of the three amplitudes
$\Db_4(\omega)$, $\Db_2(\omega_1)$, and $\Db_2(\omega_2)$. In order to
formulate this convolution correctly, we have to say a few words on
the counting of diagrams \cite{Bartels:1994jj}.

Let us return, in figure~\ref{fig:diagrams}, to the branching vertex.
Moving from the top to the bottom, it is the last interaction between
the two subsystems $(12)$ and $(34)$. Below this vertex, the system of
gluons has split into two non-interacting two-gluon states. This 'last'
vertex can be one of the $2 \to 2$ interactions inside the gluon pairs
$(23)$, $(13)$, $(24)$, or $(14)$, but not inside $(12)$ or $(34)$.
Also, it could be one of the four $2\to3$ kernels or the $2 \to 4$
kernel. Apart from that, there exists also the possibility that the
two BFKL Pomerons couple directly to the upper $R$-current 
impact factor which then provides the branching vertex.
Comparing with the integral equation for $\Db_4(\omega)$,
eq.\ (\ref{inteq4_adjoint}), we see that the diagrams summed by means of
these equations include also, as the final interactions, those inside
$(12)$ and $(34)$. For the calculation of the partial wave
$F(\omega_1,\omega_2,\omega; t_1,t_2,t)$ we therefore have to subtract
them. 
Following closely the treatment in \cite{Bartels:1994jj}, we
obtain the partial wave as the convolution
\be
\label{eq:F_convol}
\begin{split}
F(\omega_1,\omega_2, \omega ; t_1, t_2, t) =&{} \,
4\, \Db_2^{a_1a_2}(\omega_1) \otimes_{12} 
\Db_2^{a_3a_4}(\omega_2) \,\otimes_{34} 
\\
& \bigg[ \hat{\Db}_{(4;0)}^{a_1a_2a_3a_4} (\omega)
+ K^{ \{b\} \rarr \{a\} }_{2\rarr 4} \otimes \Db_2^{b_1b_2} (\omega) 
\\
&{} \hspace{0.4cm} 
+ \sum K^{ \{b\} \rarr \{a\} }_{2\rarr 3} \otimes \Db_3^{b_1b_2b_3} (\omega) 
+ {\sum}' K^{ \{b\} \rarr \{a\} }_{2\rarr 2} 
\otimes \Db_4^{b_1b_2b_3b_4} (\omega)
\bigg]
\,,
\end{split}
\ee
where the prime on the sum over the $2 \to 2$ transitions in the last
line indicates that interactions inside the gluon-pairs (12) and (34)
are not included. $\hat{\Db}_{(4;0)}^{a_1a_2a_3a_4}(\omega)$ is the
Mellin transform of the unintegrated four-gluon impact factor. The
latter appears if we do not have $s$-channel gluons. As long as we
have one or more $s$-channel gluons that contribute to the
discontinuity in $M^2$, in addition to the adjoint particle pair in
the loop, the invariant mass of the two particles in the loop is
integrated over and the integration is included in the definition of
the impact factor $\Db_{(4;0)}$ (see eq.\ (\ref{eq:def_impactn})).
Without such $s$-channel gluons the mass of the adjoint particle pair
coincides with the diffractive mass $M$ that is a fixed external
parameter. 
In this case the coupling of the four $t$-channel gluons to the quark
and scalar loop is given by an 'unintegrated' four-gluon impact factor
$D_{(4;0)}^{\text{unintegrated}}(\bk_1,\bk_2,\bk_3,\bk_4 ; M^2)$ which
carries an explicit $M^2$-dependence. It can be written as a
(unsubtracted) dispersion relation in $M^2$, and the 
discontinuity in $M^2$ entering this dispersion relation follows from the
triple discontinuities discussed in this paper. For simplicity, we
denote this discontinuity simply by
$\text{disc}_{M^2}D_{(4;0)}^{\text{unintegrated}}(\bk_1,\bk_2,\bk_3,\bk_4; M^2)
= \hat{D}_{(4;0)}(\bk_1,\bk_2,\bk_3,\bk_4; M^2)$. The Mellin transform
in $M^2$ of this unintegrated impact factor 
$\hat{D}_{(4;0)}(\bk_1,\bk_2,\bk_3,\bk_4; \omega)$ which enters the
partial wave \eqref{eq:F_convol} also follows from the
$M^2$-discontinuity, $\hat{D}_{(4;0)}(\bk_1,\bk_2,\bk_3,\bk_4 ;M^2)$.

However, apart from these two peculiarities, the second line 
of eq.\ \eqref{eq:F_convol} is nothing
but the r.h.s.\ of eq.\ (\ref{inteq4_adjoint}), the integral equation for
$\Db_4$. We therefore obtain
\begin{align}
\label{eq:F_convol1}
&F(\omega_1, \omega_2, \omega ; t_1, t_2, t) =
4\, \Db_2^{a_1a_2}(\omega_1) \otimes_{12} 
\Db_2^{a_3a_4}(\omega_2) \,\otimes_{34} \bigg[ \hat{\Db}_{(4;0)}^{a_1a_2a_3a_4}(\omega)
\notag \\
&\hspace{1cm}
+
\left( \omega - \sum_i \beta({\bf k}_i) \right) \Db_{4}^{a_1a_2a_3a_4}(\omega) 
  -  {\Db}_{(4;0)}^{a_1a_2a_3a_4} 
- \sum_{(12), (34)} K^{ \{b\} \rarr \{a\} }_{2\rarr 2} 
\otimes \Db_4^{b_1b_2b_3b_4}(\omega)\bigg] \,.
\end{align}
Using for the last term of the second line the integral equations for
$\Db_2(\omega_1)$ and $\Db_2(\omega_2)$ we finally find 
\begin{align}
 \label{eq:F_convol2}
F(\omega_1, \omega_2, \omega ; t_1, t_2, t) = &
\,4\, \Db_2^{a_1a_2}(\omega_1) \otimes_{12} 
\Db_2^{a_3a_4}(\omega_2) \otimes_{34}
\notag \\ &
\left( \hat{\Db}_{(4;0)}^{a_1a_2a_3a_4}(\omega) +( \omega - \omega_1 - \omega_2)  
\Db_{4}^{a_1a_2a_3a_4}(\omega) 
\right)
\,,
\end{align}
 where we dropped terms that do not depend on $\omega$, $\omega_1$
or $\omega_2$ and which give only a vanishing contribution to the
six-point amplitude (\ref{tripleregge}). Following now the
decomposition (\ref{split_d4_adjoint}) for $\Db_4(\omega)$, the
partial wave consists of three pieces: \be
\label{Fdecomposition}
F=F^R + F^I + F^{\text{dir}} \,.
\ee
Beginning with $F^R$ we have 
\be
\label{eq:FRfinal}
\begin{split}
F^R(\omega_1,\omega_2,\omega; t_1,t_2,t) &
\\  
= \,4\, \Db_2^{a_1a_2} (\omega_1) \otimes_{12} & \Db_2^{a_3a_4} (\omega_2)  \otimes_{34}
\left(\hat{\Db}_{(4;0)}^{Ra_1a_2a_3a_4}(\omega) 
+ ( \omega - \omega_1 - \omega_2 ) \Db_4^{R\, a_1a_2a_3a_4}(\omega) \right) \,.
\end{split}
\ee
It is possible to rewrite this in a more intuitive way. We introduce 
the 'disconnected' vertex function $V^{\text{disc}}({\lf_1},{\lf_2};\kf_1,\kf_2,\kf_3,\kf_4)$, 
\be
\label{eq:V_disc_def}
\begin{split}
V^{\text{disc}} ({\bf l}_1, {\bf l}_2; {\bf k}_1,&{\bf k}_2,{\bf k}_3, {\bf k}_4 )
\\
= 
-g^2{\bf l}_1^2 {\bf l}_2^2   
\Big[&
     \delta^{(2)}({\bf l}_1 - {\bf k}_1- {\bf k}_2- {\bf k}_3 ) 
     +
     \delta^{(2)}({\bf l}_1 - {\bf k}_1- {\bf k}_2- {\bf k}_4 ) 
     +
     \delta^{(2)}({\bf l}_1 - {\bf k}_1- {\bf k}_3- {\bf k}_4 ) 
\\     
&   
     +  
     \delta^{(2)}({\bf l}_1 - {\bf k}_2- {\bf k}_3- {\bf k}_4 ) 
    -
     \delta^{(2)}({\bf l}_1 - {\bf k}_1- {\bf k}_2 ) 
     -
     \delta^{(2)}({\bf l}_1 - {\bf k}_1- {\bf k}_3  )
\\
&     -
     \delta^{(2)}({\bf l}_1 - {\bf k}_1- {\bf k}_4)  \Big] \,,
\end{split}
\ee
and use for the color factor 
\be
\label{colfac}
C= \delta_{a_1a_2} \delta_{a_3a_4} d^{a_1a_2a_3a_4} = \frac{1}{2 N_c} (N_c^2 - 1)^2
\,,
\ee
such that we arrive at
\be
\begin{split}
  F^R(\omega_1,\omega_2,\omega; t_1,t_2,t) =& \,
4 \,C \, 
\Db_2(\omega_1) \otimes_{12} \Db_2(\omega_2)  \otimes_{34} 
\\ &
\left(
\hat{\Db}_{(4;0)}^{Ra_1a_2a_3a_4}(\omega) 
+ \left( \omega - \omega_1 - \omega_2 \right)V^{\text{disc}} \otimes   \Db_2(\omega)   \right) \,
\end{split} 
\ee
For $F^I$ we have 
\be
\label{FIfinal}
\begin{split}
F^I(\omega_1,\omega_2,\omega; t_1,t_2,t)=& \,4\, \Db_2^{a_1a_2}
(\omega_1) \otimes_{12}
\Db_2^{a_3a_4} (\omega_2)  \otimes_{34}
\\
&
\left(V_{2 \rightarrow 4}^{a_1a_2a_3a_4} \Db_2 (\omega) +{\sum}' K^{ \{b\}
    \rarr \{a\} }_{2\rarr 2} \otimes \Db_4^{I\,b_1b_2b_3b_4}(\omega) \right) \,,
\end{split}
\ee 
where the sum in the last term extends over the pairs (13),
(14), (23), and (24) and we made use of the integral equation for
$\Db_4^I(\omega)$ (see eq.\ (\ref{inteq4tilde_adjoint})) and for
$\Db_2(\omega_1)$ and $\Db_2(\omega_2)$. In a similar way we obtain
for $F^{\rm dir}$ 
\be
\label{Fdirfinal}
\begin{split}
F^{{\rm dir}}(\omega_1,\omega_2,\omega; t_1,t_2,t)=& \,
4 \,\Db_2^{a_1a_2} (\omega_1) \otimes_{12}
\Db_2^{a_3a_4} (\omega_2)  \otimes_{34}
\\
&\left( \hat{\Db}_{(4;0)\text{dir}}^{a_1a_2a_3a_4}(\omega) + {\sum}'
  K^{ \{b\} \rarr \{a\} }_{2\rarr 2} \otimes \Db_{4\,
    \text{dir}}^{b_1b_2b_3b_4}(\omega)  \right) \,.  
\end{split}
\ee
The structure of the
unintegrated four-gluon impact factor with 
fermions in the fundamental representation of $SU(N_c)$ in the
loop, for the case $t=t_1=t_2=0$, has been given in
\cite{Bartels:1994jj}. For our analysis we define
\begin{align}
  {\hat \Db}^{a_1a_2a_3a_4}_{(4;0)}
=
 {\hat \Db}^{Ra_1a_2a_3a_4}_{(4;0)}
+
 {\hat \Db}^{a_1a_2a_3a_4}_{(4;0)\text{dir}}
\end{align}
with
\begin{align}
   {\hat \Db}^{Ra_1a_2a_3a_4}_{(4;0)} =  
d^{a_1a_2a_3a_4}\left({\hat \Db}_{S,(4;0)}
+{\hat \Db}_{F,(4;0)} \right) \,,
\end{align}
and
\begin{align}
 {\hat \Db}^{a_1a_2a_3a_4}_{(4;0)\text{dir}} = 
\frac{1}{2N_c}(\delta_{a_1a_2}\delta_{a_3a_4}
+\delta_{a_1a_3}\delta_{a_2a_4}+\delta_{a_1a_4}\delta_{a_2a_3})
\left({\hat \Db}_{S,(4;0)}+{\hat \Db}_{F,(4;0)} \right) \,. 
\end{align}
The results are 
\be
\begin{split}
\hat\Db_{F,{(4;0)}}^{hh'}\hspace*{-0.8cm}&\hspace*{0.8cm}
(\bk,-\bk,\bk',-\bk';M^2) \\
=& {}
\,\frac{g^4N_c}{2}\, \frac{M^2}{(2\pi)^3}
\int_0^1 d\alpha\int d^2 \bl\, \alpha(\alpha-1)\delta(\alpha(1-\alpha)M^2-\bl^2)\\
&\left[ -(2\alpha-1)^2 \beps^{(h)}\cdot
\left(\frac{\bl+\bk}{D(\bl+\bk)}+\frac{\bl-\bk}{D(\bl-\bk)}-2 \frac{\bl}{D(\bl)}\right)\right.
\\
& \hspace{0.3cm}\times
\left(\frac{\bl+\bk'}{D(\bl+\bk')}+\frac{\bl-\bk'}{D(\bl-\bk')}
-2 \frac{\bl}{D(\bl)}\right)\cdot\beps^{(h')}
\\
&\hspace{0.3cm}+\beps^{(h')}\cdot\left(\frac{\bl+\bk}{D(\bl+\bk)}
+\frac{\bl-\bk}{D(\bl-\bk)}-2 \frac{\bl}{D(\bl)}\right)
\left(\frac{\bl+\bk'}{D(\bl+\bk')}+\frac{\bl-\bk'}{D(\bl-\bk')}
-2 \frac{\bl}{D(\bl)}\right)\cdot\beps^{(h)}
\\
&\hspace{0.3cm}- \left.\beps^{(h)}\cdot\beps^{(h')}\left(\frac{\bl+\bk}{D(\bl+\bk)}
+\frac{\bl-\bk}{D(\bl-\bk)}-2 \frac{\bl}{D(\bl)}\right)
\cdot\left(\frac{\bl+\bk'}{D(\bl+\bk')}+\frac{\bl-\bk'}{D(\bl-\bk')}
-2 \frac{\bl}{D(\bl)}\right) 
\right]
\end{split} 
\ee 
for transversely polarized $R$-currents, and 
\be
\begin{split}
\hat\Db_{F,{(4;0)}}^{LL} &(\bk,-\bk,\bk',-\bk';M^2)  \\
=& - 2g^4N_c\, Q^2 \frac{M^2}{(2\pi)^3}\int_0^1 
d\alpha\int d^2 \bl\, \alpha^3(\alpha-1)^3\delta(\alpha(1-\alpha)M^2-\bl^2) \\
& \times\left(\frac{1}{D(\bl+\bk)}+\frac{1}{D(\bl-\bk)}-\frac{2}{D(\bl)}\right)\cdot
\left(\frac{1}{D(\bl+\bk')}+\frac{1}{D(\bl-\bk')}-\frac{2}{D(\bl)}\right)
\end{split}
\ee
for longitudinally polarized $R$-currents. 
The denominators in these expressions are 
\be
    D(\bk) = \alpha (1-\alpha) Q^2 + \bk^2 \,.
\ee
Similarly, the scalar contributions to the unintegrated four-gluon impact factor are
\be
\begin{split}
\hat\Db_{S,{(4;0)}}^{hh'}&(\bk,-\bk,\bk',-\bk';M^2) \\
=&{} 
\,2  g^4 N_c\, \frac{M^2}{(2\pi)^3}\int_0^1 d\alpha\int d^2 \bl\, 
\alpha^2 (\alpha-1)^2 \delta(\alpha(1-\alpha)M^2-\bl^2)
\\
&\times\beps^{(h)}\cdot\left(\frac{\bl+\bk}{D(\bl+\bk)}+\frac{\bl-\bk}{D(\bl-\bk)}
-2 \frac{\bl}{D(\bl)}\right)
\left(\frac{\bl+\bk'}{D(\bl+\bk')}+\frac{\bl-\bk'}{D(\bl-\bk')}
-2 \frac{\bl}{D(\bl)}\right)\cdot\beps^{(h')}\\
\end{split}
\ee
and
\be
\begin{split}
\hat\Db_{S,{(4;0)}}^{LL} & (\bk,-\bk,\bk',-\bk';M^2) \\
=&{}
\,2g^4  N_c\,  Q^2\frac{M^2}{(2\pi)^3} \int_0^1 d\alpha\int d^2 \bl\, 
(\alpha-1/2)^2\alpha^2(\alpha-1)^2 \delta(\alpha(1-\alpha)M^2-\bl^2)\\
&\times\left(\frac{1}{D(\bl+\bk)}+\frac{1}{D(\bl-\bk)}-\frac{2}{D(\bl)}\right)
\left(\frac{1}{D(\bl+\bk')}+\frac{1}{D(\bl-\bk')}-\frac{2}{D(\bl)}\right) \,.
\end{split}
\ee

In the forward case $t = t_1 = t_2 = 0$, 
$\Db_{2}(\omega_1) =\Db_{2}({\bf k},-{\bf k}, \omega_1)=\Db_{2}({\bf k}^2,\omega_1)$ 
and  
$\Db_{2}(\omega_2) =\Db_{2}({\bf k}',-{\bf k}',\omega_2)=\Db_{2}({{\bf k}'}^2,\omega_2)$, 
i.\,e.\ the complete dependence on the azimuthal angle of the momenta ${\bf k}$ 
and ${{\bf k}'}$ is inside the unintegrated impact factors. 
After integration over the angles of the momenta $\bl$, $\bk$, and $\bk'$, 
the sum of fermionic and scalar contributions simplifies. The results are
\begin{align}
  \hat\Db_{(4;0)}^{hh'}&(\bk^2,{\bk'}^2;M^2) =
\notag \\
 =&  \int \frac{d \varphi_\bk }{2\pi} \int \frac{d \varphi_{\bk'} }{2\pi} 
\left[
\hat\Db_{F(4;0)}^{hh'}(\bk,-\bk,\bk',-\bk';M^2)  
+
\hat\Db_{S(4;0)}^{hh'}(\bk,-\bk,\bk',-\bk';M^2)
\right] 
\notag \\
=&{} \,
\frac{g^4N_c}{4}  \delta_{hh'} 
\int_0^1 
d \alpha\,I_v(\bk^2,\alpha,M^2) I_v({\bk'}^2,\alpha,M^2)
\end{align}
and
\begin{align}
  \hat\Db_{(4;0)}^{LL}&(\bk^2,{\bk'}^2;M^2) =
\notag \\
 =&  \int \frac{d \varphi_\bk }{2\pi} \int \frac{d \varphi_{\bk'} }{2\pi} 
\left[
\hat\Db_{F(4;0)}^{LL}(\bk,-\bk,\bk',-\bk';M^2)  
+
\hat\Db_{S(4;0)}^{}(\bk,-\bk,\bk',-\bk';M^2)
\right] 
\notag \\
=&{}\,
\frac{g^4N_c}{4}  \int_0^1 d \alpha\, \alpha^2 (1-\alpha)^2 
I_s(\bk^2,\alpha,M^2) I_s({\bk'}^2,\alpha,M^2) \,.
\end{align}
Here we have defined 
\be
\begin{split}
\frac{\bl}{\bl^2} I_v(\bk^2,\alpha,M^2)&=
\int_0^{2\pi} \frac{d\varphi}{2\pi} \left( \frac{\bl+\bk}{D(\bl+\bk)}
+\frac{\bl-\bk}{D(\bl-\bk)}-2 \frac{\bl}{D(\bl)}\right)\\
&= \frac{\bl}{\bl^2} \left( \frac{Q^2 - M^2}{Q^2 + M^2} - 
\frac{\bk^2 +\alpha (1-\alpha) (Q^2 - M^2)}
{\sqrt{(\bk^2+\alpha(1-\alpha) (Q^2 - M^2))^2 + 4 \alpha^2(1-\alpha)^2 M^2 Q^2}}\right)
\end{split}   
\ee
and 
\be
\begin{split}
I_s (\bk^2,& \alpha,M^2)=  \sqrt{M^2 Q^2}
\int_0^{2\pi} \frac{d\varphi_\bk}{2\pi} \left(\frac{1}{D(\bl+\bk)}
+\frac{1}{D(\bl-\bk)}-\frac{2}{D(\bl)}\right) \\
&= 2 \left( \frac{ \sqrt{M^2 Q^2}}{\sqrt{(\bk^2+\alpha(1-\alpha) (Q^2 - M^2))^2 
+ 4 \alpha^2(1-\alpha)^2 M^2 Q^2}} 
- \frac{ \sqrt{M^2 Q^2}}{\alpha(1-\alpha)(Q^2 + M^2)}\right) ,
\end{split}
\ee
and $\varphi_\bk$ ($\varphi_{\bk'}$) denotes the angle of the vector $\bk$ ($\bk'$), while 
the $\delta$-function has been used to set $\bl^2 = \alpha(1-\alpha) M^2$. 

We expect that the results found above will be useful for a detailed 
comparison with similar amplitudes calculated on the supergravity side of 
the AdS/CFT correspondence \cite{Bartels:2009jb}. 

Let us finally comment on the large-$N_c$ limit. Beginning with the
first term, $F^R$, we observe in eq.\ (\ref{eq:FRfinal}) that the color
factor $C$ goes as $N_c^3/2$, see \eqref{colfac}. It may also be useful to note that,
making use of the M\"obius representation of the BFKL amplitude, in
eq.\ (\ref{eq:V_disc_def}) only the last two terms contribute. In our
result for $F^I$, on the r.h.s.\ of eq.\ (\ref{FIfinal}) the terms with the
$2\to2$ kernels are color suppressed, and as a result the two BFKL
amplitudes $\Db_2(\omega_1)$ and $\Db_2(\omega_2)$ are directly
attached to the triple Pomeron vertex. Finally, for $F^{\text{dir}}$
in eq.\ (\ref{Fdirfinal}), again the terms with the $2\to2$ kernels are
color suppressed, and the BFKL Pomerons couple directly to the new
piece in the impact factor, $\Db_{(4;0)\, \text{dir}}$.

Diagrammatically, the three pieces $F^R$, $F^I$, and $F^{\text{dir}}$
are illustrated in figure~\ref{fig:largeNc}. As a result of the large-$N_c$ 
limit, the BKP four-gluon states which had been present for
finite $N_c$ have disappeared: in order to find such states for large
$N_c$ it would be necessary to go to higher order $R$-current
correlators.%
\FIGURE{
\includegraphics[width=10cm]{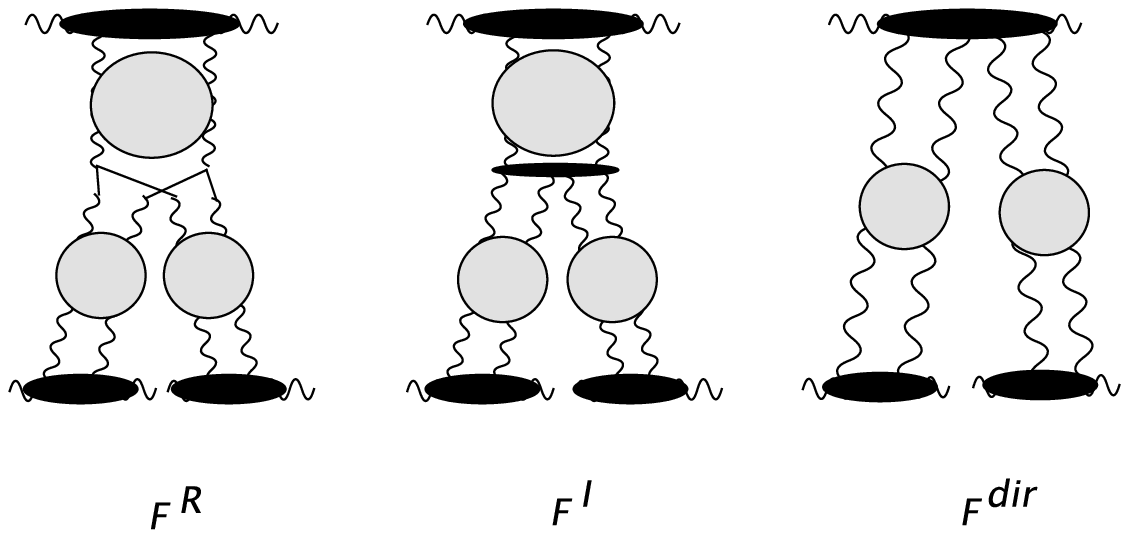}
\caption{Large-$N_c$ limit of the six-point function \label{fig:largeNc}}
}

Compared to non-supersymmetric $\text{SU}(N_c)$ gauge theories with fundamental
quarks, the most striking difference is the presence of the last piece
which exists only in the supersymmetric extension where all particles
are in the adjoint representation. The triple Pomeron vertex, on the
other hand, is the same in both cases.
A detailed discussion of the topological expansion of the triple Pomeron 
vertex in the amplitudes above and its large-$N_c$ behavior 
has been given in \cite{Bartels:2009ms,Bartels:2009zc}. 

\section{Summary and outlook}
\label{sec:outlook}

In this paper we have studied, in the generalized leading logarithmic 
approximation, the high energy behavior of ${\cal N}=4$ SYM in the 
triple Regge limit. It is this kinematic regime which, in QCD, exhibits the 
M\"obius invariant triple Pomeron vertex. As the main result, we have found 
that in ${\cal N}=4$ SYM, with the fermions and scalars belonging to the adjoint 
representation of the gauge group $\text{SU}(N_c)$, the four-gluon impact factor 
contains a novel piece whose existence can be 
traced back to the adjoint representation of the fermions and scalars. 
It has no counterpart in QCD where the quarks transform in the fundamental 
representation of the gauge group. In the six-point amplitude, this 
additional piece in the impact factor generates a coupling of the 
four-gluon state to the external currents which is absent in QCD.
On the other hand, the triple Pomeron vertex in ${\cal N}=4$ SYM, in leading 
order, is the same as in the non-supersymmetric case. This supports the 
fundamental nature of this building block of Reggeon field theory: because of 
Regge factorization it has to be independent of the coupling to the external 
projectiles. In our case, this coupling is mediated by the 
impact factors in which the difference between ${\cal N}=4$ SYM and 
non-supersymmetric QCD is manifest: the fact that in both cases the 
triple Pomeron vertex is the same proves that factorization is indeed 
satisfied. 

\section*{Acknowledgements}

We are grateful to M.\ Salvadore, V.\ Schomerus, and G.\,P.\ Vacca 
for helpful discussions. J.\,B.\ thanks L.\ Yaffe for useful conversations. 
C.\,E.\ would like to thank G.\ Korchemsky and G.\ Moore for useful discussions. 
J.\,B.\ and C.\,E.\ thank the Galileo Galilei Institute for Theoretical
Physics for the hospitality and the INFN for partial support during 
a stay in Firenze where part of this work was carried out. 
C.\,E.\ would like to thank the II. Institut 
f\"ur Theoretische Physik of the University of Hamburg for hospitality. 
M.\,H.\ would like to thank the Paul Scherrer Institut Villigen 
and the Instituto de F\'isica Corpuscular Valencia for hospitality. 
The work of C.\,E.\ was supported by the Alliance Program of the
Helmholtz Association (HA216/EMMI). M.\,H.\ thanks the 
DFG Graduiertenkolleg `Zuk\"unftige Ent\-wicklungen in der 
Teilchenphysik' and DESY Hamburg for financial support. 
\begin{appendix}

\section{Color algebra}
\label{sec:color}

In the calculations in section \ref{sec:external} and in appendix \ref{appfive} 
we make use of the following $\text{SU}(N_c)$ color identities, 
\bea
\label{contr2f}
f_{lak} f_{kbl} 
&=& {} 
- N_c \,\delta_{ab} \,, 
\\
\label{contr3f}
f_{kal} f_{lbm} f_{mck} 
&=& {}
- \frac{N_c}{2} \,f_{abc}  \,,
\\
\label{contr4f}
f_{kal} f_{lbm} f_{mcn}f_{ndk} 
&=& {}
N_c\, d^{abcd} + \frac{1}{2} 
 ( \delta_{ab}\delta_{cd} + \delta_{ac}\delta_{bd} 
 + \delta_{ad}\delta_{bc} ) \,,
\\
\label{contr5f}
f_{kal} f_{lbm} f_{mcn}f_{ndo}f_{oek} 
&=& {} 
N_c \, f^{abcde} \nn \\
&&{}
+ \frac{1}{4} ( \delta_{ab}f_{cde} 
+ \delta_{ac}f_{bde} + \delta_{ad}f_{bce} 
+ \delta_{ae}f_{bcd} \nn \\
&& {}
\hspace{0.7cm}
+ f_{ade}\delta_{bc} + f_{ace}\delta_{bd} 
+ f_{acd}\delta_{be} + f_{abe}\delta_{cd}\nn \\
&&{}
 \hspace{0.7cm}
+ f_{abd}\delta_{ce} + f_{abc}\delta_{de})
\,,
\eea
where the tensors $d^{abcd}$ and $f^{abcde}$ are defined in 
(\ref{d_abcd}) and (\ref{f_abcde}), respectively. 
The first three identities are well-known while the last one has been 
derived in \cite{Bartels:1999aw}. That paper also contains a description 
of a general and convenient way to obtain such identities using 
birdtrack notation. 

It is worth pointing out that the relevant color tensors for the adjoint 
impact factors in the Pomeron channel satisfy 
\be
\label{zweimaltrace}
\tr (T^{a_1} \dots T^{a_n}) + (-1)^n \tr (T^{a_n} \dots T^{a_1}) = 
2 \,\tr (T^{a_1} \dots T^{a_n}) 
\,.
\ee

We finally note that the $d$-tensor of \eqref{d_abcd} can be decomposed 
according to 
\be
\label{decomposedtens}
 d^{a b c d} =\frac{1}{2 N_c} \delta_{ab} \delta_{cd} 
                   + \frac{1}{4} (d_{abk}d_{kcd} - f_{abk}f_{kcd} ) 
\,.
\ee

\section{The five-gluon amplitude} 
\label{appfive}

In this appendix we would like to investigate the five-gluon 
amplitude in ${\cal N}=4$ SYM in the EGLLA. The 
main focus of this paper has been the four-gluon amplitude 
relevant for the six-point $R$-current correlator. 
The considerations in the present and in the following 
appendix \ref{appsix} aim at a better understanding of the field theory 
structure of the amplitudes in the EGLLA. The main properties 
of that structure and its relation to reggeization have been described 
in section \ref{sec:amplitudeseglla} for the case of the four-gluon 
amplitude. Here we want to discuss the five-gluon amplitude. We 
will show that, as a consequence of reggeization, it can be written 
completely in terms of elements of the 2-dimensional effective 
field theory which have been found already in the four-gluon amplitude. 

Let us start with the five-gluon impact factor $\Db_{(5;0)}$. 
I consists again of a fermionic and a scalar contribution. 
We follow the same steps as in section \ref{sec:external}. 
In the case of five gluons the color tensors relevant for the impact 
factor in the fundamental representation (i.\,e.\ in QCD) are of the type 
\be
\label{f_abcde}
f^{abcde} = \frac{1}{i}[\tr(t^a t^b t^c t^d t^e) - \tr(t^e t^d t^c t^bt^a)] 
\,.
\ee
For the case of an impact factor consisting of a loop made of particles 
in the adjoint representation one obtains instead 
\begin{eqnarray}
\label{f_abcde:adj}
\lefteqn{
\frac{1}{i}[\tr(T^a T^b T^c T^d T^e) - \tr(T^e T^d T^c T^b T^a)] 
}
\nn \\
&=&{} 2N_c f^{abcde} 
+ \frac{1}{2} ( \delta_{ab}  f_{cde} + \delta_{ac} f_{bde} + \delta_{ad} f_{bce} 
 + \delta_{ae} f_{bcd} + \delta_{bc} f_{ade} + \delta_{bd} f_{ace} 
\notag \\
&&{}\hspace*{2.4cm}
+ \delta_{be} f_{acd} + \delta_{cd} f_{abe} + \delta_{ce} f_{abd} + \delta_{de} f_{abc}
) 
\,,
\end{eqnarray}
where we have used \eqref{contr5f} and \eqref{zweimaltrace}. 
Again, an additional color tensor structure 
occurs which was not present in the fundamental representation. 
For the fermionic contribution $\Db_{F,(5;0)}$ to the 
impact factor this implies that we can decompose it 
into a part that is a multiple of the fundamental impact factor and an 
additional term as 
\begin{eqnarray}
\label{eq:D_50:adj}
\lefteqn{
\Db_{F,(5;0)}^{a_1 a_2 a_3 a_4 a_5} (\kf_1, \kf_2, \kf_3, \kf_4, \kf_5)
\notag } \\ 
&=&{}
2 N_c R\,
D_{(5;0)}^{a_1 a_2 a_3 a_4 a_5} (\kf_1, \kf_2, \kf_3, \kf_4, \kf_5) 
+ \Db_{F,(5;0) \,\text{dir}}^{ a_1 a_2 a_3 a_4 a_5} 
(\kf_1, \kf_2, \kf_3, \kf_4, \kf_5) 
\,.
\end{eqnarray}
With the help of the explicit expression for $D_{(5;0)}$ found in 
\cite{Bartels:1999aw}, 
\bea
\label{d50}
\lefteqn{D_{(5;0)}^{a_1a_2a_3a_4a_5}(\kf_1,\kf_2,\kf_3,\kf_4,\kf_5)}
\\
&=&{}  - g^3 \{ f^{a_1a_2a_3a_4a_5} \, [
                  D_{(2;0)}(1234,5) + D_{(2;0)}(1,2345) - D_{(2;0)}(15,234)]
\nn \\
&&{}  \,
              + f^{a_2a_1a_3a_4a_5} \, [
                  D_{(2;0)}(1345,2) - D_{(2;0)}(12,345)
                  + D_{(2;0)}(125,34) - D_{(2;0)}(134,25) ]
\nn \\
  &&{}  \,
              + f^{a_1a_2a_3a_5a_4} \, [
                  D_{(2;0)}(1235,4) - D_{(2;0)}(14,235)
                  + D_{(2;0)}(145,23) - D_{(2;0)}(123,45) ]
\nn \\
&&{}  \,
              + f^{a_1a_2a_4a_5a_3} \, [
                  D_{(2;0)}(1245,3) - D_{(2;0)}(13,245)
                  + D_{(2;0)}(135,24)
- D_{(2;0)}(124,35) ] \}
\,, 
\nn
\eea
we can thus write 
\begin{eqnarray}
\label{eq:D_50:adj_expl}
\lefteqn{
\Db_{F,(5;0)}^{a_1 a_2 a_3 a_4 a_5} (\kf_1, \kf_2, \kf_3, \kf_4, \kf_5)
\notag } \\ 
&=&{}
- g^3 \{ f^{a_1a_2a_3a_4a_5} \, [
                  \Db_{F,(2;0)}(1234,5) + \Db_{F(2;0)}(1,2345) - \Db_{F,(2;0)}(15,234)]
\nn \\
&&{}  \,
              + f^{a_2a_1a_3a_4a_5} \, [
                  \Db_{F,(2;0)}(1345,2) - \Db_{F,(2;0)}(12,345)
                  + \Db_{F,(2;0)}(125,34) - \Db_{F,(2;0)}(134,25) ]
\nn \\
  &&{}  \,
              + f^{a_1a_2a_3a_5a_4} \, [
                  \Db_{F,(2;0)}(1235,4) - \Db_{F,(2;0)}(14,235)
                  + \Db_{F,(2;0)}(145,23) - \Db_{F,(2;0)}(123,45) ]
\nn \\
&&{}  \,
              + f^{a_1a_2a_4a_5a_3} \, [
                  \Db_{F,(2;0)}(1245,3) - \Db_{F,(2;0)}(13,245)
                  + \Db_{F,(2;0)}(135,24)
- \Db_{F,(2;0)}(124,35) ] \}
\nn \\
&&{}+ \Db_{F,(5;0) \,\text{dir}}^{ a_1 a_2 a_3 a_4 a_5} 
(\kf_1, \kf_2, \kf_3, \kf_4, \kf_5) 
\,
\end{eqnarray}
The additional piece, $\Db_{F,(5;0) \,\text{dir}}$,
is obtained explicitly by replacing the 
different tensors of type $f^{abcde}$ in eq.\ (\ref{d50}) by the 
additional color tensors emerging in eq.\ (\ref{f_abcde:adj}). 
A lengthy but straightforward calculation shows that the result can 
also be expressed in terms of the additional piece 
$\Db_{(4;0)\; \text{dir}}$ found previously in the four-gluon 
amplitude (see eq.\ \eqref{D40_adj}) as 
\begin{eqnarray}
  \label{eq:D50add_byD40addd}
\lefteqn{
\Db^{ a_1 a_2 a_3 a_4 a_5}_{F,(5;0)\,\text{dir}}
(\kf_1, \kf_2, \kf_3, \kf_4, \kf_5)} 
\nn \\
&=&{}\frac{g}{2} 
\left[
f_{a_1a_2c} \,\Db_{F,(4;0) \text{dir}}^{ca_3a_4a_5}(12,3,4,5) 
+ f_{a_1a_3c} \,\Db_{F,(4;0)\; \text{dir}}^{ca_2a_4a_5}(13,2,4,5) \right. 
\nn \\
&&{} \hspace{1.2cm} 
+ \,f_{a_1a_4c} \,\Db_{F,(4;0)\text{dir}}^{ca_2a_3a_5}(14,2,3,5) 
+ f_{a_1a_5c} \,\Db_{F,(4;0)\,\text{dir}}^{ca_2a_3a_4}(15,2,3,4) 
\nn \\
&&{} \hspace{1.2cm} 
+ \,f_{a_2a_3c} \,\Db_{F,(4;0)\,\text{dir}}^{a_1ca_4a_5}(1,23,4,5) 
+ f_{a_2a_4c} \,\Db_{F,(4;0)\,\text{dir}}^{a_1ca_3a_5}(1,24,3,5) 
\nn \\
&&{} \hspace{1.2cm} 
+\, f_{a_2a_5c} \,\Db_{F,(4;0)\,\text{dir}}^{a_1ca_3a_4}(1,25,3,4) 
+ f_{a_3a_4c} \,\Db_{F,(4;0)\,\text{dir}}^{a_1a_2ca_5}(1,2,34,5) 
\nn \\
&&{} \hspace{1.2cm} \left.
+ \,f_{a_3a_5c} \,\Db_{F,(4;0)\,\text{dir}}^{a_1a_2ca_4}(1,2,35,4) 
+ f_{a_4a_5c} \,\Db_{F,(4;0)\,\text{dir}}^{a_1a_2a_3c}(1,2,3,45)
\right]
\,.
\end{eqnarray}

It is straightforward to show that equations analogous to 
(\ref{eq:D_50:adj_expl}) and (\ref{eq:D50add_byD40addd}) are valid 
also for the scalar contribution $\Db_{S,(5;0)}$ to the impact factor, 
that is also the additional piece $\Db_{S,(5;0)\,\text{dir}}$ can 
be expressed in terms of the additional piece $\Db_{S,(4;0)\,\text{dir}}$ 
of the four-gluon amplitude. 
Consequently, we obtain relations for the full impact factor 
$\Db_{(5;0)}=\Db_{F,(5;0)} + \Db_{S,(5;0)}$ which are completely 
analogous to (\ref{eq:D_50:adj_expl}) and (\ref{eq:D50add_byD40addd}). 
More precisely, eq.\ (\ref{eq:D_50:adj_expl}) and eq.\ (\ref{eq:D50add_byD40addd}) 
are valid after dropping the index $F$ in all terms. The second of 
these relations has been given explicitly for the full impact factor 
in eq.\ \eqref{eq:D50add_byD40add}. 

This representation shows that the additional piece $\Db_{(5;0)\, \text{dir}}$ 
again exhibits reggeization. In each term in eq.\ \eqref{eq:D50add_byD40add} 
a pair of gluons in the color octet representation acts as a single gluon 
which enters the amplitude $\Db_{(4;0)\, \text{dir}}$. 
In a certain sense this gluon can be regarded as a composite object of 
the two gluons merging into it. 
The full expression for $\Db_{(5;0)\, \text{dir}}$ is then obtained by 
summing over all possible pairs of gluons. We recall that the amplitude 
$\Db_{(4;0)\, \text{dir}}$ is fully symmetric in its momentum and 
color arguments such that it is not relevant at which position the 
more composite gluon formed from the pair is inserted in 
$\Db_{(4;0)\, \text{dir}}$.

We can now put the picture derived form the three- and four-gluon 
amplitudes to the test by considering the integral equation for the 
five-gluon amplitude. According to the expected field theory 
structure we should find that the five-gluon amplitude $\Db_5$ 
reggeizes, that means it should be possible to express it 
completely in terms of elements that are already present 
in the lower amplitudes. This should now in particular 
include $\Db_{(4;0)\, \text{dir}}$ which we have identified 
as a new element of our field theory.
The evolution equation for $\Db_5^{a_1a_2a_3a_4a_5}$ reads 
\bea
\label{inteq5_adjoint}
\left(\omega - \sum_{i=1}^5 \beta(\kf_i) \right) 
\Db_5^{a_1a_2a_3a_4a_5} 
&=& {}
\Db_{(5;0)}^{a_1a_2a_3a_4a_5} 
+ K^{ \{b\} \rarr \{a\} }_{2\rarr 5} \otimes \Db_2^{b_1b_2} \nn \\
&&{}+ \sum K^{ \{b\} \rarr \{a\} }_{2\rarr 4} \otimes \Db_3^{b_1b_2b_3} 
+ \sum K^{ \{b\} \rarr \{a\} }_{2\rarr 3} \otimes \Db_4^{b_1b_2b_3b_4} \nn \\
&&{}+ \sum K^{ \{b\} \rarr \{a\} }_{2\rarr 2} \otimes \Db_5^{b_1b_2b_3b_4b_5}
\,, 
\eea
which can be graphically illustrated as 
\bea
\label{inteqD5diag}
\left( \omega - \sum_{i=1}^5 \beta(\kf_i)\right) 
\eqeps{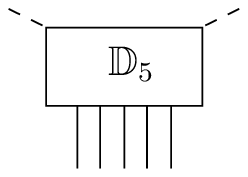}{-2} 
&=&{} 
\eqeps{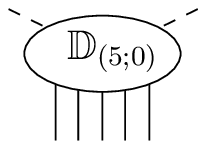}{-1} 
+ \eqeps{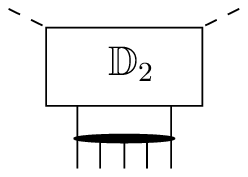}{-2} 
+ \sum \eqeps{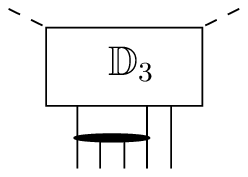}{-2} 
\nn \\
&&{}
+ \sum \eqeps{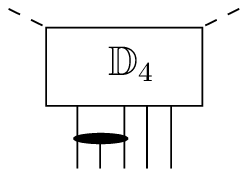}{-2} 
+ \sum \eqeps{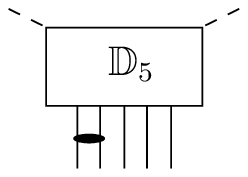}{-2} 
\,.
\eea
In view of the results found for the amplitudes with up to four 
gluons, and since the adjoint impact factor with five gluons 
\eqref{eq:D_50:adj} consists of a multiple of the fundamental 
impact factor and a new (additional) piece, 
it is natural to decompose the full amplitude $\Db_5$ according to 
\be
\label{D5adjdecomp}
\Db_5 = \Db_5^R + \Db_5^I + \Db_{5\, \text{dir}}
\,,
\ee
where the first two parts are multiples of the corresponding 
amplitudes $D_5^R$ and $D_5^I$ known from QCD with 
fundamental quarks. There, these two parts exhaust the full 
amplitude, whereas here we will have an additional part 
$\Db_{5\, \text{dir}}$. According to this picture we find, 
using the results of \cite{Bartels:1999aw}, 
\bea
\label{d5r_adjoint}
% PLEASE do not change the layout of this equation
\lefteqn{
\Db_5^{R\,a_1a_2a_3a_4a_5}(\kf_1,\kf_2,\kf_3,\kf_4,\kf_5)= } 
\\
  &=&{}  - g^3 \{ f^{a_1a_2a_3a_4a_5} \, [ 
                  \Db_2(1234,5) + \Db_2(1,2345) - \Db_2(15,234)] 
   \nn \\
  & &{}  \phantom{ - g^3 } 
              + f^{a_2a_1a_3a_4a_5} \, [ 
                  \Db_2(1345,2) - \Db_2(12,345)
                  + \Db_2(125,34) - \Db_2(134,25) ]
   \nn \\
  & &{}  \phantom{ - g^3 } 
              + f^{a_1a_2a_3a_5a_4} \, [ 
                  \Db_2(1235,4) - \Db_2(14,235)
                  + \Db_2(145,23) - \Db_2(123,45) ]
   \nn \\
  & &{}  \phantom{ - g^3 } 
              + f^{a_1a_2a_4a_5a_3} \, [ 
                  \Db_2(1245,3) - \Db_2(13,245)
                  + \Db_2(135,24) 
\nn 
- \Db_2(124,35) ] \} 
\nn
\eea
and 
\bea
\label{d5isolution_adjoint}
%
% PLEASE do not change the layout of this equation.
%
\lefteqn{
\Db_5^{I\,a_1a_2a_3a_4a_5}(\kf_1,\kf_2,\kf_3,\kf_4,\kf_5) 
}
\nn \\
&&{}=\frac{g}{2} \left\{ f_{a_1a_2c} \Db_4^{I\,ca_3a_4a_5}(12,3,4,5) 
+ f_{a_1a_3c} \Db_4^{I\,ca_2a_4a_5}(13,2,4,5) \right. \nn \\
&&{} \hspace{.5cm} 
+ \,f_{a_1a_4c} \Db_4^{I\,ca_2a_3a_5}(14,2,3,5) 
+ f_{a_1a_5c} \Db_4^{I\,ca_2a_3a_4}(15,2,3,4) \nn \\
&&{} \hspace{.5cm} 
+ \,f_{a_2a_3c} \Db_4^{I\,a_1ca_4a_5}(1,23,4,5) 
+ f_{a_2a_4c} \Db_4^{I\,a_1ca_3a_5}(1,24,3,5) \nn \\
&&{} \hspace{.5cm} 
+\, f_{a_2a_5c} \Db_4^{I\,a_1ca_3a_4}(1,25,3,4) 
+ f_{a_3a_4c} \Db_4^{I\,a_1a_2ca_5}(1,2,34,5) \nn \\
&&{} \hspace{.5cm} \left.
+ \,f_{a_3a_5c} \Db_4^{I\,a_1a_2ca_4}(1,2,35,4) 
+ f_{a_4a_5c} \Db_4^{I\,a_1a_2a_3c}(1,2,3,45)
\right\}
\,.
\eea
We have directly written these two parts in terms of the 
adjoint amplitudes $\Db_2$. 
For the fermionic part we have in particular $\Db_{F,5}^R=2 N_c R D_5^R$ 
and $\Db_{F,5}^I = 2 N_c R D_5^I$. Inserting then 
eq.\ \eqref{D5adjdecomp} into the integral equation 
\eqref{inteq5_adjoint} and invoking the integral equation 
for $D_5$ treated in \cite{Bartels:1999aw} we obtain 
a new integral equation for the additional part 
$\Db_{5\,\text{dir}}$, 
\bea
\label{inteq5add_adjoint}
\left(\omega - \sum_{i=1}^5 \beta(\kf_i) \right) 
\Db_{5\,\text{dir}}^{a_1a_2a_3a_4a_5} 
&=&{}
\Db_{(5;0)\,\text{dir}}^{a_1a_2a_3a_4a_5} 
+ \sum K^{ \{b\} \rarr \{a\} }_{2\rarr 3} \otimes 
\Db_{4\,\text{dir}}^{b_1b_2b_3b_4} 
\nn \\
&&{}
+ \sum K^{ \{b\} \rarr \{a\} }_{2\rarr 2} \otimes 
\Db_{5\,\text{dir}}^{b_1b_2b_3b_4b_5}
\,. 
\eea
We recall that the inhomogeneous term $\Db_{(5;0)\,\text{dir}}$ 
reggeizes and is a superposition of amplitudes 
$\Db_{(4;0)\,\text{dir}}$ of the form \eqref{eq:D50add_byD40add}. 
Therefore the structure of this equation coincides with that of the 
integral equations for $\Db_3$ and $D_5^I$. It can be 
solved employing a known identity involving the 
integral kernels $K_{2 \to 2}$ and $K_{2 \to 3}$. 
In terms of the diagrams of \eqref{diagkernel2m} 
this identity can be written as 
\begin{eqnarray}
\label{3gluonidentity} 
\picb{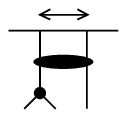} + \picb{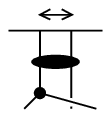} 
+ \picb{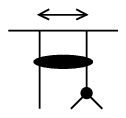} &=&
 \frac{2}{g}\,\picb{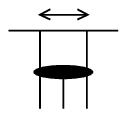} + \picb{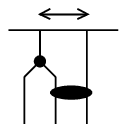}
+ \picb{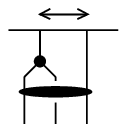}
\nonumber \\
&& + \picb{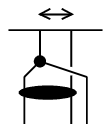}+ \picb{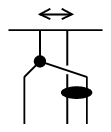}
+ \picb{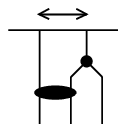}+ \picb{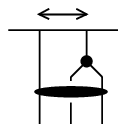}
\,,
\label{reggeizebilder}
\end{eqnarray}
where the unaffected gluons are not drawn. The vertices in which two 
lines merge into one, drawn here with a small full dot, indicate that the 
two lower gluons merge into the upper one with a color factor $f_{abc}$ 
and in such a way that their transverse momenta are added. 
The arrows on the top of the diagrams indicate that the identity holds 
for gluons in amplitudes that 
are symmetric under the exchange of this pair of gluons (or symmetric 
under the exchange of any pair of gluons). This is the case for the 
amplitude $\Db_{4\,\text{dir}}$ as we have discussed above. 
For a more detailed description of the application of the identity 
\eqref{3gluonidentity} to completely analogous situations we refer 
the reader to \cite{Bartels:1999aw} or \cite{Braunewell:2005ct}. 
In this way one finds the solution of eq.\ \eqref{inteq5add_adjoint} to 
have the form 
\bea
\label{d5add_solution_adjoint}
%
% PLEASE do not change the layout of this equation.
%
\lefteqn{\Db_{5\,\text{dir}}^{a_1a_2a_3a_4a_5}(\kf_1,\kf_2,\kf_3,\kf_4,\kf_5) }
\nn \\
&&{}=\frac{g}{2} \left\{ f_{a_1a_2c} \Db_{4\,\text{dir}}^{ca_3a_4a_5}(12,3,4,5) 
+ f_{a_1a_3c} \Db_{4\,\text{dir}}^{ca_2a_4a_5}(13,2,4,5) \right. \nn \\
&&{} \hspace{.5cm} 
+ \,f_{a_1a_4c} \Db_{4\,\text{dir}}^{ca_2a_3a_5}(14,2,3,5) 
+ f_{a_1a_5c} \Db_{4\,\text{dir}}^{ca_2a_3a_4}(15,2,3,4) \nn \\
&&{} \hspace{.5cm} 
+ \,f_{a_2a_3c} \Db_{4\,\text{dir}}^{a_1ca_4a_5}(1,23,4,5) 
+ f_{a_2a_4c} \Db_{4\,\text{dir}}^{a_1ca_3a_5}(1,24,3,5) \nn \\
&&{} \hspace{.5cm} 
+\, f_{a_2a_5c} \Db_{4\,\text{dir}}^{a_1ca_3a_4}(1,25,3,4) 
+ f_{a_3a_4c} \Db_{4\,\text{dir}}^{a_1a_2ca_5}(1,2,34,5) \nn \\
&&{} \hspace{.5cm} \left.
+ \,f_{a_3a_5c} \Db_{4\,\text{dir}}^{a_1a_2ca_4}(1,2,35,4) 
+ f_{a_4a_5c} \Db_{4\,\text{dir}}^{a_1a_2a_3c}(1,2,3,45)
\right\}
\,.
\eea
One immediately recognizes that this amplitudes again exhibits 
reggeization. It is a superposition of the four-gluon amplitudes 
$\Db_{4\,\text{dir}}$, and in each term of the superposition 
two gluons in a color octet combine to form a more composite 
reggeized gluon that enters the amplitude $\Db_{4\,\text{dir}}$. 
The sum over all possible pairs of this kind gives the additional 
part of the full five-gluon amplitude $\Db_5$. One can easily 
check from eq.\ \eqref{d5add_solution_adjoint} that this expression 
satisfies the Ward-type identities of \cite{Ewerz:2001fb}, and 
hence all parts of the amplitude $\Db_5$ share that property. 

In summary, we have solved the integral equation 
\eqref{inteq5_adjoint} for the five-gluon amplitude $\Db_5$ 
and have expressed the solution in terms of the lower amplitudes 
$\Db_2$, $\Db_4^I$ and $\Db_{4\,\text{dir}}$. Diagrammatically, 
the solution has the form 
\be
\label{Db5soldiag}
\Db_5 = \sum \eqepsres{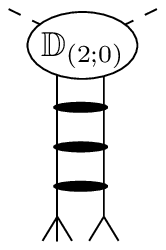}{-2}{1.8cm} 
+ \sum \eqepsres{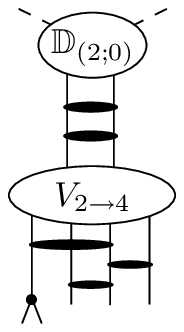}{-2}{2cm} + \sum \eqepsres{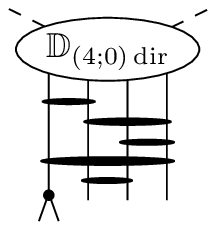}{-2}{2.3cm} 
\,.
\ee
The first two terms are multiples of the amplitudes known from 
the case of QCD with fundamental quarks. The last terms is an additional 
term that exhibits exactly the features expected from the picture 
of the effective field theory structure discussed above, in particular 
it reggeizes exactly as expected and fulfills the Ward-type identities 
of \cite{Ewerz:2001fb}. Hence we find that the five-gluon amplitude 
precisely meets our expectations based on that underlying picture. 

\section{Some remarks on the six-gluon amplitude}
\label{appsix}

In this appendix we want to consider some aspects of the six-gluon 
amplitude $\Db_6$ in the EGLLA for ${\cal N}=$ SYM. 
Our motivation is that it would obviously be interesting to see 
whether and how the field theory structure 
found above in the amplitudes $\Db_n$ for$n\le 5$ continues in the amplitudes 
with more than five gluons. Already in QCD with fundamental quarks 
the six-gluon amplitude involves some additional complication, see the 
discussion in \cite{Bartels:1999aw}. A full analysis of the six-gluon 
is beyond the scope of the present paper and is left for future work. 
It might be particularly interesting to find out whether another 
new element of the effective field theory emerges that would 
couple six gluons directly to the adjoint quark loop -- in analogy 
to the element $\Db_{(4;0)\, \text{add}}$ with four gluons found 
above. From the results obtained here the picture emerges that in general 
(that is for arbitrary $n$) the amplitudes $\Db_n$ 
always contain a multiple of the amplitudes $D_n$ with fundamental 
quarks plus additional new terms. Those latter contain on the one 
hand terms originating from the scalar contribution to the impact 
factor, and their structure concerning reggeization is very similar to 
that observed in the terms coming from the fermionic impact factor. 
On the other hand, there are new terms, for example the one coupling 
four gluons directly to the impact factor. 

Although we do not make an attempt here to solve the whole six-gluon 
amplitude we can rather easily make an interesting 
observation concerning $\Db_6$ which in fact confirms 
that expectation. Namely, one finds that similarly to the case 
of the lower amplitudes the terms known 
from the corresponding amplitude $D_6$ of QCD with fundamental 
fermions are reproduced (again up to a normalization factor) 
together with their scalar counterparts, 
and additional terms are generated. In particular, the same 
transition from two to six gluons via the vertex $V_{2 \to 6}$ 
computed in \cite{Bartels:1999aw} (containing both a 
coupling of a Pomeron to two Odderons and a transition from 
one to three Pomerons) occurs again in the amplitude $\Db_6$. 

The diagrams with six gluons coupling to the fermion or scalar loop 
give rise to a color tensor of the type 
\bea
\label{6adjtens}
\tr (T^a T^b T^c T^d T^e T^f) + \tr (T^f T^e T^d T^c T^b T^a) 
&=&{} 2 \, \tr (T^a T^b T^c T^d T^e T^f) 
\nn \\
&=&{} - 2 f_{kal} f_{lbm} f_{mcn}f_{ndo}f_{oep}f_{pfk}
\eea
instead of the tensor 
\be
\label{6fundtens}
d^{abcdef}= \tr (t^a t^b t^c t^d t^e t^f) + \tr (t^f t^e t^d t^c t^b t^a) 
\ee
obtained for quarks in the fundamental representation. 
We can therefore obtain the adjoint fermion impact factor 
$\Db_{F,(6;0)}$ from the known expression for the 
fundamental impact factor $D_{(6;0)}$ (see \cite{Bartels:1999aw}), 
by replacing the color tensors of type \eqref{6fundtens} 
in this expression by the corresponding color tensors of 
type \eqref{6adjtens}, besides decorating 
it with the usual factor $R$ accounting for the different 
numbers of fermionic degrees of freedom running around the loop. 
But now one has \cite{Bartels:1999aw} 
\bea
\label{6fdecomp}
f_{kal} f_{lbm} f_{mcn}f_{ndo}f_{oep}f_{pfk} &=& {}
- N_c d^{abcdef} 
\nn 
\\
&&{} -\frac{1}{2} ( \delta_{ab} d^{cdef}
+ \delta_{ac} d^{bdef} + \delta_{ad} d^{bcef} 
+ \delta_{ae} d^{bcdf} \nn \\
&&{} \hspace{.7cm}
+ \delta_{af} d^{bcde} + d^{adef} \delta_{bc} 
+ d^{acef} \delta_{bd} + d^{acdf} \delta_{be} \nn \\
&&{} \hspace{.7cm}
+ d^{acde} \delta_{bf} + d^{abef} \delta_{cd} 
+ d^{abdf} \delta_{ce} + d^{abde} \delta_{cf} \nn \\
&&{} \hspace{.7cm}
+ d^{abcf} \delta_{de} + d^{abce} \delta_{df} 
+ d^{abcd} \delta_{ef} )
\nn \\
&&{} + \frac{1}{8} [ 
(d_{abc} d_{def} + f_{abc} f_{def}) 
+ (d_{abd} d_{cef} + f_{abd} f_{cef}) \nn \\
&&{} \hspace{.7cm}
+ (d_{abe} d_{cdf} + f_{abe} f_{cdf})
+ (d_{abf} d_{cde} + f_{abf} f_{cde})\nn \\
&&{} \hspace{.7cm}
+ (d_{acd} d_{bef} + f_{acd} f_{bef})
+ (d_{ace} d_{bdf} + f_{ace} f_{bdf})\nn \\
&&{} \hspace{.7cm}
+ (d_{acf} d_{bde} + f_{acf} f_{bde})
+ (d_{ade} d_{bcf} + f_{ade} f_{bcf})\nn \\
&&{} \hspace{.7cm}
+ (d_{adf} d_{bce} + f_{adf} f_{bce})
+ (d_{aef} d_{bcd} + f_{aef} f_{bcd})
]
\\
\label{thetatensdef}
&=&{} - N_c d^{abcdef} + \Theta^{abcdef}
\,,
\eea
where the last equation defines the tensor $\Theta^{abcdef}$. 
Therefore inserting the color tensors of type \eqref{6adjtens} 
one naturally obtains two contributions to the fermion 
impact factor $\Db_{F,(6;0)}$, 
\bea
\label{Db60mitadd}
\Db_{F,(6;0)}^{a_1a_2a_3a_4a_5a_6} (\kf_1,\kf_2,\kf_3,\kf_4,\kf_5,\kf_6)
&=&{} 
2 N_c R D_{(6;0)}^{a_1a_2a_3a_4a_5a_6}
   (\kf_1,\kf_2,\kf_3,\kf_4,\kf_5,\kf_6)
\nn \\
&&{}+ 
\Db^{a_1a_2a_3a_4a_5a_6}_{(6;0)\,\text{dir}} (\kf_1, \kf_2, \kf_3, \kf_4,\kf_5,\kf_6)
\,
\eea
where
\bea
%
% PLEASE do not change the layout of this equation.
%
\label{d60}
\lefteqn{D_{(6;0)}^{a_1a_2a_3a_4a_5a_6}
   (\kf_1,\kf_2,\kf_3,\kf_4,\kf_5,\kf_6)= } \nn \\
&=&  g^4 \{ d^{a_1a_2a_3a_4a_5a_6} \,[ 
  D_{(2;0)}(12345,6) + D_{(2;0)}(1,23456) - D_{(2;0)}(16,2345) ] \nn \\
& &\,+\,d^{a_2a_1a_3a_4a_5a_6} \,[ 
  D_{(2;0)}(13456,2) - D_{(2;0)}(1345,26) 
+ D_{(2;0)}(126,345) \nn \\
&&\hspace{3cm} -  D_{(2;0)}(12,3456) ] \nn \\
& &\, +\,d^{a_1a_2a_3a_4a_6a_5} \,[ 
 D_{(2;0)}(12346,5) - D_{(2;0)}(1234,56) 
+ D_{(2;0)}(156,234) \nn \\
&&\hspace{3cm} - D_{(2;0)}(15,2346) ] \nn \\
& &\, +\,d^{a_2a_1a_3a_4a_6a_5} \,[
 - D_{(2;0)}(1256,34) - D_{(2;0)}(1346,25) 
 + D_{(2;0)}(125,346) \nn \\
&& \hspace{3cm} + D_{(2;0)}(134,256) ] \nn \\
& &\, +\,d^{a_3a_1a_2a_4a_5a_6} \,[
 D_{(2;0)}(12456,3) - D_{(2;0)}(1245,36) 
 + D_{(2;0)}(136,245) \nn \\
&&\hspace{3cm} - D_{(2;0)}(13,2456) ] \nn \\
& &\,+\,d^{a_1a_2a_3a_5a_6a_4} \,[
 D_{(2;0)}(12356,4) - D_{(2;0)}(1235,46) 
+ D_{(2;0)}(146,235) \nn \\
&&\hspace{3cm} - D_{(2;0)}(14,2356) ] \nn \\
& &\, +\,d^{a_2a_1a_3a_5a_6a_4} \,[
- D_{(2;0)}(1246,35) - D_{(2;0)}(1356,24)
+ D_{(2;0)}(124,356) \nn \\
&&\hspace{3cm} + D_{(2;0)}(135,246) ] \nn \\
& &\, +\,d^{a_1a_2a_3a_6a_5a_4} \,[
- D_{(2;0)}(1236,45) - D_{(2;0)}(1456,23) 
+ D_{(2;0)}(123,456) \nn \\
& & \hspace{3cm} + D_{(2;0)}(145,236) ] 
\}
\eea
Consequently, the first term on the r.h.s.\ of eq.\ \eqref{Db60mitadd} 
can again be expressed as a sum of two-gluon impact factors 
$\Db_{F,(2;0)}= 2N_c R D_{(2;0)}$. 
A similar expression can be obtained for the scalar impact factor
$\Db_{S,(6;0)}$.

The first contribution in eq.\ (\ref{Db60mitadd}), arising from the
first term in the decomposition eq.\ \eqref{thetatensdef} of the
adjoint color tensors, is just a multiple of the known six-gluon
impact factor in the fundamental representation. The second
contribution arises from the $\Theta$-tensors in 
\eqref{thetatensdef} (as defined in (\ref{6fdecomp}) when
inserting \eqref{6adjtens} for the $d$-tensors in \eqref{d60}). 
It would be straightforward to write out the new
additional term explicitly. As we have pointed out before it would be
interesting to study in detail its relation to the new element
$\Db_{F,(4;0)\,\text{dir}}$ found there. This, however, is beyond the
scope of the present paper. Here we restrict ourselves to the
important finding that also the adjoint impact factor with six gluons
is similar to those with four and five gluons: it reproduces as one of
its contributions a multiple of the fundamental impact factor and in
addition contains a new part that occurs only if the fermions are
taken in the adjoint representation. A similar additional contribution
is obtained from the adjoint scalars in the impact factor.

Inserting this result for $\Db_{F,(6;0)\,\text{dir}}$ (and the analogous 
scalar part) in the integral
equation for the full amplitude $\Db_6$ we easily recognize that also
the full amplitude consists of two parts, one of which is a multiple
of the full six-gluon amplitude $D_6$ with fundamental quarks (together 
with its scalar counterpart) and
hence inherits the conformal field theory structure found in that
amplitude. In particular, this part contains exactly the same transition vertex
from two to six gluons found in \cite{Bartels:1999aw}. The structure
of the other part remains to be investigated.

\end{appendix}

\end{document}